\documentclass[prb, twocolumn, showpacs, superscriptaddress]{revtex4}
\pdfoutput=1 
\usepackage{graphicx}

\begin{document}

\title{Modulation of Electrical Conduction Through Individual Molecules on 
Silicon 
by the Electrostatic Fields of Nearby Polar Molecules: Theory and Experiment}

\author{George Kirczenow}

\affiliation{Department of Physics, Simon Fraser
University, Burnaby, British Columbia, Canada V5A 1S6}

\affiliation{The Canadian Institute
for Advanced Research, Nanoelectronics Program, Canada.}

\author{Paul G. Piva}

\altaffiliation{Present Address: Institute for National Measurement Standards, 
NRC,
Ottawa, Ontario, Canada K1A 0R6.}

\affiliation{National Institute for Nanotechnology,
National Research Council of Canada, Edmonton, Alberta
T6G 2V4, Canada, and Department of Physics, University of
Alberta, Edmonton, Alberta T6G 2J1, Canada}

\author{Robert A. Wolkow}

\affiliation{National Institute for Nanotechnology,
National Research Council of Canada, Edmonton, Alberta
T6G 2V4, Canada, and Department of Physics, University of
Alberta, Edmonton, Alberta T6G 2J1, Canada}

\affiliation{The Canadian Institute
for Advanced Research, Nanoelectronics Program, Canada.}

\date{\today}


\begin{abstract}

We report on the synthesis, scanning tunneling microscopy (STM) and
theoretical modeling of the electrostatic and transport properties of
one-dimensional organic heterostructures consisting of contiguous lines of 
CF$_3$-- and OCH$_3$--substituted styrene molecules on silicon.
The electrostatic fields emanating from these polar molecules are
found, under appropriate conditions,
to strongly influence electrical conduction through nearby molecules 
and the underlying substrate. For suitable alignment of the OCH$_3$ groups
of the OCH$_3$--styrene molecules in the molecular chain, 
their combined electric fields are
shown by {\em ab initio} density functional calculations to give rise
to potential profiles along the OCH$_3$--styrene chain that result
in strongly enhanced conduction through OCH$_3$--styrene molecules near
the heterojunction for moderately low negative substrate bias, as is observed
experimentally. Under similar bias conditions, dipoles associated with the
 CF$_3$ groups are found in both experiment and in theory to depress 
 transport in the underlying silicon. Under positive substrate bias, simulations 
suggest that the differing structural and electrostatic properties of the CF3-styrene 
molecules may lead to a more sharply localized conduction enhancement 
near the heterojunction at low temperatures.
Thus choice of substituents, their 
attachment site on the host styrene molecules on silicon and the orientations
of the molecular dipole and higher multipole moments  
provide a means of differentially tuning transport on the molecular 
scale.     

\end{abstract}

\maketitle

\section{Introduction}
\label{Intro}

During the last decade a great deal of research 
has focussed  on electrical conduction through individual molecules\cite
{Ratner06, Tao06, GK}.
Many molecules are electrically polarized due to chemical charge transfer 
between unlike atoms. This results in 
electric fields that should influence electrical conduction 
through
such molecules and through molecules in their vicinity.  However, there have 
been
few direct experimental investigations of such effects in the context of 
molecular-scale nanoelectronics. 
Recently experimental studies of the effects of charged
chemical species attached to a molecule on electrical conduction through the 
{\em same} molecule have
been reported \cite{Venk2ring,Venkataraman}. The presence of a charged 
dangling bond on a silicon
surface has been observed to affect electrical conduction through nearby 
molecules\cite{Nature} 
and, conversely, transport through adjacent silicon atoms has been found
to be perturbed by dipole fields due to molecules located
elsewhere in the Si 7$\times$7 cell\cite{Harikumar}.  
However, no experimental work
directly probing the effects of electric fields emanating from 
polar molecules on electrical 
conduction through {\em other} individual molecules has been reported to 
date. 
This topic
is explored experimentally and theoretically in the present 
article.\cite{brief} The
influence of these electric fields on electrical conduction through the
underlying semiconductor substrate to which the molecules are bound
is also examined.\cite{brief}

The specific systems that we study here are one-dimensional (1D) organic 
heterostructures consisting of 
chains of substituted styrene molecules grown on hydrogen-terminated Si 
substrates by 
the self-directed growth mechanism
described by Lopinski and coworkers\cite{Lopinski} originally for styrene on 
H:Si(100) 2$\times$1 but subsequently applied to a wide range of molecules. 
The substituents in the present work are CF$_3$ and OCH$_3$ groups, one 
of which
replaces a single hydrogen atom bound to the aromatic carbon ring of each 
styrene molecule. In each case the substituent is in the ``para" position.
Each heterostructure consists of a line of surface bound 
OCH$_3$--styrene molecules  joined 
end-to-end (at the heterojunction)
to a line of CF$_3$--styrene molecules.
The OCH$_3$ group donates electrons to the aromatic ring of the styrene and 
is therefore positively charged
 while the CF$_3$ group withdraws electrons and is negatively charged. 
However, electric
 polarization also occurs {\em within} the CF$_3$ and OCH$_3$ groups 
themselves 
 and we find the resulting molecular  multipole fields to also play an important 
role in
electrical conduction in these systems. 
 
In the experimental work presented here electrical conduction between the 
silicon substrate and a tungsten scanning
tunneling microscope (STM) tip via the molecular heterostructures was 
measured at room temperature in ultra-high vacuum under a variety
of bias conditions. Strongly enhanced conduction was observed 
through a group of several 
OCH$_3$--styrene molecules near the heterojunction at low and moderate 
negative substrate bias. Under positive substrate bias similar 
pronounced enhancement of the  
molecular conduction near CF$_3$--styrene/OCH$_3$--styrene 
heterojunctions was absent. In related heterostructures, CF$_3$--styrene 
lines in a {\em side-by-side} configuration were found to locally depress transport 
originating from filled states in the underlying silicon.   As will be explained 
below these 
qualitative differences can be understood as arising from the different 
structural, electronic and electrostatic properties of OCH$_3$--styrene and 
CF$_3$--styrene molecules on silicon.

The OCH$_3$--styrene/CF$_3$--styrene heterowires were modeled 
theoretically by determining their ground state electrostatic potential profiles
(at zero applied bias) by means of {\em ab initio} density functional 
theory-based computations, and then
employing semi-empirical tight binding models together with the {\em ab 
initio} electrostatic profiles, solution of the Lippmann-Schwinger equation and 
Landauer theory to calculate the electric current between the tungsten STM tip 
and the silicon substrate under bias via the molecular heterostructure. We 
note that while the use of density functional theory for performing {\em 
transport} calculations in molecular wires lacks a fundamental justification and 
is increasingly being questioned in the literature,\cite
{Delaney,Sai,Toher05,Darancet,Evers,Ke07,Pemmaraju,Prodan,offsets,
Koentoppreview,Thygesen78,Toher78,Thygesen8,
GK} the use of density functional theory to calculate ground state electrostatic 
potentials is justified (at least in principle) by a lemma proved originally by 
Hohenberg and Kohn. \cite{HK} The usefulness
of such density functional theory-based {\em electrostatic potential} 
calculations  for the present system is supported also by the
fact that the present theory is able to explain the experimentally observed 
phenomena outlined in the preceding paragraph.

The
theoretical work that is presented here demonstrates that electric fields due to 
intramolecular charge transfer within
OCH$_3$--styrene and CF$_3$--styrene molecules can (for appropriate 
molecular geometries)
result in enhanced electrical conduction
through certain molecules in chains of OCH$_3$--styrene and 
CF$_3$--styrene molecules on silicon,
consistent with the experimental STM data: These electric fields shift the 
energies of the molecular HOMO and LUMO states of
{\em like} molecules in different parts of the chain by {\em differing} 
amounts with the 
result that resonant conduction begins at lower
bias voltages on some molecules in the chain than on others. The differences 
in the observed behavior under positive and negative applied bias and 
between the OCH$_3$--styrene and CF$_3$--styrene molecules are 
accounted for in terms of this electrostatically modulated resonant transport 
mechanism. 

The present theory predicts that conduction through individual molecules in 
such systems can
be changed by orders of magnitude by varying the conformations of (and 
hence the arrangement of charges in)
molecules in their vicinity. 
This raises the possibility of molecular switches of a new type that depend for
their operation not on conformational changes in the molecular wire carrying 
the current (as has been discussed
previously\cite{switch1,switch,switch2}) but on conformational changes in 
{\em other} nearby molecules that
result in changes in the energy level structure of the molecular wire. An 
advantage of a conformational
switch based on this different principle is that the molecular wire that carries 
the current need not be a
moving part in the device. The shorter range of electrostatic fields due to 
molecular
dipoles (and higher multipoles) than those of electric monopoles such as the 
charged dangling bond
studied in Ref. \onlinecite{Nature} is also advantageous, making it possible in 
principle to achieve higher
densities of molecular switches controlled by molecular dipoles in potential 
device applications. The greater chemical stability of a polar
molecule than a charged radical is also an important advantage. Furthermore 
manipulating the
orientations of molecular dipoles may be an attractive alternative\cite{dipole} 
to inducing switching by
charging and discharging atomic or molecular-scale constituents of the 
ultimate nanoscale electronic devices.  

This article is organized as follows: The experimental methodology and 
results are presented in Section \ref{Expt}. The theoretical model is explained 
and justified in Section \ref{Model}. 
A discussion of the relevant energy level ordering in these systems is 
presented in Section \ref{Level Ordering}. 
The different structures that the 
molecules may assume when bound to silicon are described in Section \ref
{Geom}.
Our
theoretical results for a particular conformation of a 
OCH$_3$--styrene/CF$_3$--styrene molecular chain on silicon 
are presented in Section \ref{Case} for positive and negative substrate bias, 
together with some comments as to how these theoretical results 
may relate to the experimental data.  
In Section \ref{Relationship} we clarify the relationship 
between structure,
electric potentials and transport  in the OCH$_3$--styrene/CF$_3$--styrene 
heterostructures by 
considering systematically other examples of possible molecular chain 
geometries. Theoretical results demonstrating  
that the current enhancement near the OCH$_3$--styrene/CF$_3$--styrene 
junction is specifically an electrostatic effect are reported
in Section \ref{EvE}. Simulations of
heterostructures that include single and triple rows of CF$_3$--styrene 
molecules and of the influence of these molecules 
on the electrostatic potentials in the underlying
silicon and on electron transport are reported in Section
\ref{Single-Triple}.
Further discussion of the relationship between the theory
and experiment is presented
in Section \ref{conclusions}.  

\section{Experiment}
\label{Expt}

STM experiments were performed under vacuum on hydrogen-terminated Si
(100) 2$\times$1 surfaces.  Samples 
were cleaved from arsenic-doped (resistivity $<$ 0.005 
$\Omega$cm) Si (100) oriented wafers,
and mounted into molybdenum sample holders.  Samples were loadlocked 
into a vacuum system 
(background pressure
$<$ 1$\times$10$^{-10}$ Torr) and degassed at $700\,^{\circ}\mathrm{C}$ for 
8 hours.  The samples were
flash annealed to $1250\,^{\circ}\mathrm{C}$ to remove the surface oxide and 
re-order the crystal
surfaces.  During annealing, heating was temporarily suspended if the system 
pressure exceeded
4$\times$10$^{-10}$ Torr.  Clean crystalline surfaces were routinely produced 
with defect
densities below 5\%.

After cool down and inspection in the STM, the silicon crystals were 
transferred to the preparation chamber
for hydrogen termination.  Molecular hydrogen was leaked into the system (1$
\times$10$^{-6}$ Torr) and a hot
tungsten filament (${\sim}1600\,^{\circ}\mathrm{C}$) positioned 10 cm from 
silicon sample
($\mathrm{T}=300\,^{\circ}\mathrm{C}$) dissociated the molecular gas into 
reactive atomic
hydrogen.  A 13 minute exposure produced samples with a quasi-saturated 2
$\times$1 silicon monohydride
surface.\cite{Boland}
 
One dimensional molecular organic heterowires were grown 
under vacuum using the 
self-directed growth mechanism
reported by Lopinski and coworkers\cite{Lopinski} for styrene on H:Si(100) 2$
\times$1.  For styrene
(consisting of an aromatic carbon ring bound to a vinyl group), self-directed 
growth results from a chain
reaction between the vinyl group of the styrene and a surface dangling bond 
(i.e. silicon radical exposed by a
missing hydrogen atom on the surface).  In forming a chemical bond with the 
exposed silicon atom, the terminal
carbon atom on the vinyl group breaks one of its double bonds to the adjacent 
carbon atom, leaving the
unsatisfied bonding electron on the adjacent carbon atom to abstract a 
hydrogen atom from an
adjacent Si dimer.  The newly formed dangling bond is then free to repeat the
process by reacting with another styrene molecule.  Multiple reactions lead to 
well ordered one dimensional
chains of styrene bound along a given side of a silicon dimer row.
\cite{Lopinski}

This self-directed growth mode has been observed to
occur for various alkene and carbonyl containing molecules 
including functionalized styrenes\cite{DiLabio,Tong,Kruse,Piva}
making the
formation of heterowires a simple matter of sequentially dosing the H:silicon 
surface with the desired chemical
precursors.\cite{caveat} The fortuitous alignment of the aromatic rings in these 
structures make them ideally
suited for probing transport effects resulting from overlapping $\pi$ and $\pi$* 
states along the molecular
chain axis.  The introduction of a chemical discontinuity at the heterojunction 
provides a means of isolating
and studying effects resulting from specific intermolecular interactions.  
Electron donating and withdrawing
substituent groups are of particular interest as they modify the energy 
alignment and spatial distribution of
$\pi$ and $\pi$* states in host aromatic molecules.  These effects lead to 
systematic variations in reaction
rates\cite{Hammett} and ionisation potentials\cite{DiLabio99,Anagaw} in 
substituted aromatic compounds. 
Corresponding effects on single molecule transport are under current 
investigation.\cite{Venkataraman}

Substituent effects in STM imaging of 4-methylstyrene/styrene heterowires 
were reported in Reference
\onlinecite{stymeth}.  While the substituted methyl group had a strong (bias 
dependent) influence on the differential molecular
height and corrugation resolved on either side of the heterojunction, the 
imaging
characteristics of the heterowires could be understood qualitatively without 
considering
the effects of intermolecular electrostatic interactions on the electronic 
structure of the heterowires; features reflecting such interactions were not 
observed in the experimental data.  The present work revisits the earlier 
heterowire
line-growth experiments with more strongly perturbing substituent groups:\cite
{DiLabio99,Anagaw} 4-methoxystyrene
(OCH$_3$--styrene), and 4-trifluoromethylstyrene (CF$_3$--styrene).

\begin{figure*}[b]
\includegraphics[width=1.00\linewidth, clip=true, trim=0.0 350.0 0.0 100.0]{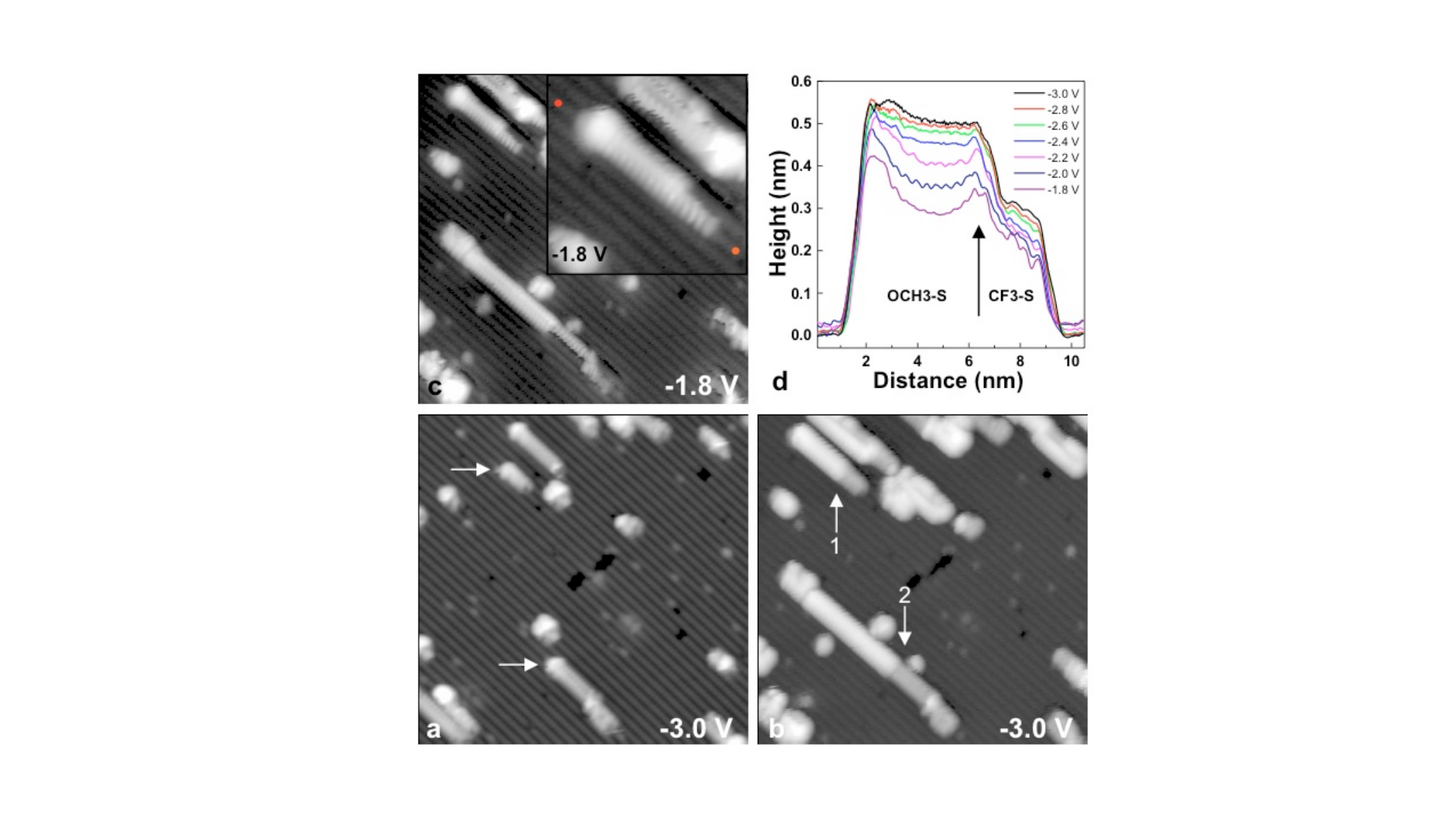}
\caption{\label{Fig_1expt}Color online. Constant current filled-state STM 
images showing 
the growth of two CF$_3$--styrene/OCH$_3$--styrene
heterowires.  (a) $V_s = -3.0$V.  H:Si(100) surface following a 10L exposure 
of CF$_3$--styrene.  Arrows
indicate positions of chemically reactive dangling bonds at the ends of the 
CF$_3$--styrene line segments.  (b)
Following a 10L exposure of OCH$_3$--styrene, the two CF$_3$--styrene line 
segments in (a) have been extended to
form two CF$_3$--styrene/OCH$_3$--styrene heterowires (`1' and `2').  At $V_s 
= -3.0$V, the OCH$_3$ images
higher (brighter) than the CF$_3$--styrene segments.  (c) $V_s = -1.8$V.  The 
OCH$_3$--styrene near the
molecular heterojunctions in heterowires 1 and 2 image with enhanced 
height.  Molecules at the end of the  OCH$_3$--styrene line segment in 
heterowire 1 also
image with enhanced height in response to the terminal dangling bond.  Inset:  
Close-up of heterowire 1. 
Molecules are bound to the right-hand side of the dimer row indicated by the 
red dots.  (d) Constant-current
topographic cross-sections extracted from bias dependent imaging of 
heterowire 1 along the trench to the right
of the attachment dimers in (c).  The height envelope for the heterostructure 
extends between $1$ nm and
$9.5$ nm along the abscissa.  The height maxima associated with the 
terminal dangling bond, and the molecular heterojunction are
at
$2.3$ nm, and $6.4$ nm, respectively.  At elevated bias (-3 V), the 
OCH$_3$--styrene images with
approximately uniform height from beyond the terminal dangling bond to the 
heterojunction.  As the bias decreases
in magnitude, the OCH$_3$--styrene images with decreased height as the 
molecular $\pi$ states drop below the
tip Fermi-level.  At -1.8 V, localised height enhancement in the 
OCH$_3$--styrene due to the terminal dangling
bond and also near the molecular heterojunction (black arrow) is most 
evident.  Image areas
(a)-(c): 26nm $\times$ 26nm.  Inset area: 8.5nm $\times$8.5 nm.  Tunnel 
current: 40 pA.}
\end{figure*}

\begin{figure*}[b]
\includegraphics[width=1.00\linewidth, clip=true, trim=0.0 200.0 0.0 100.0]{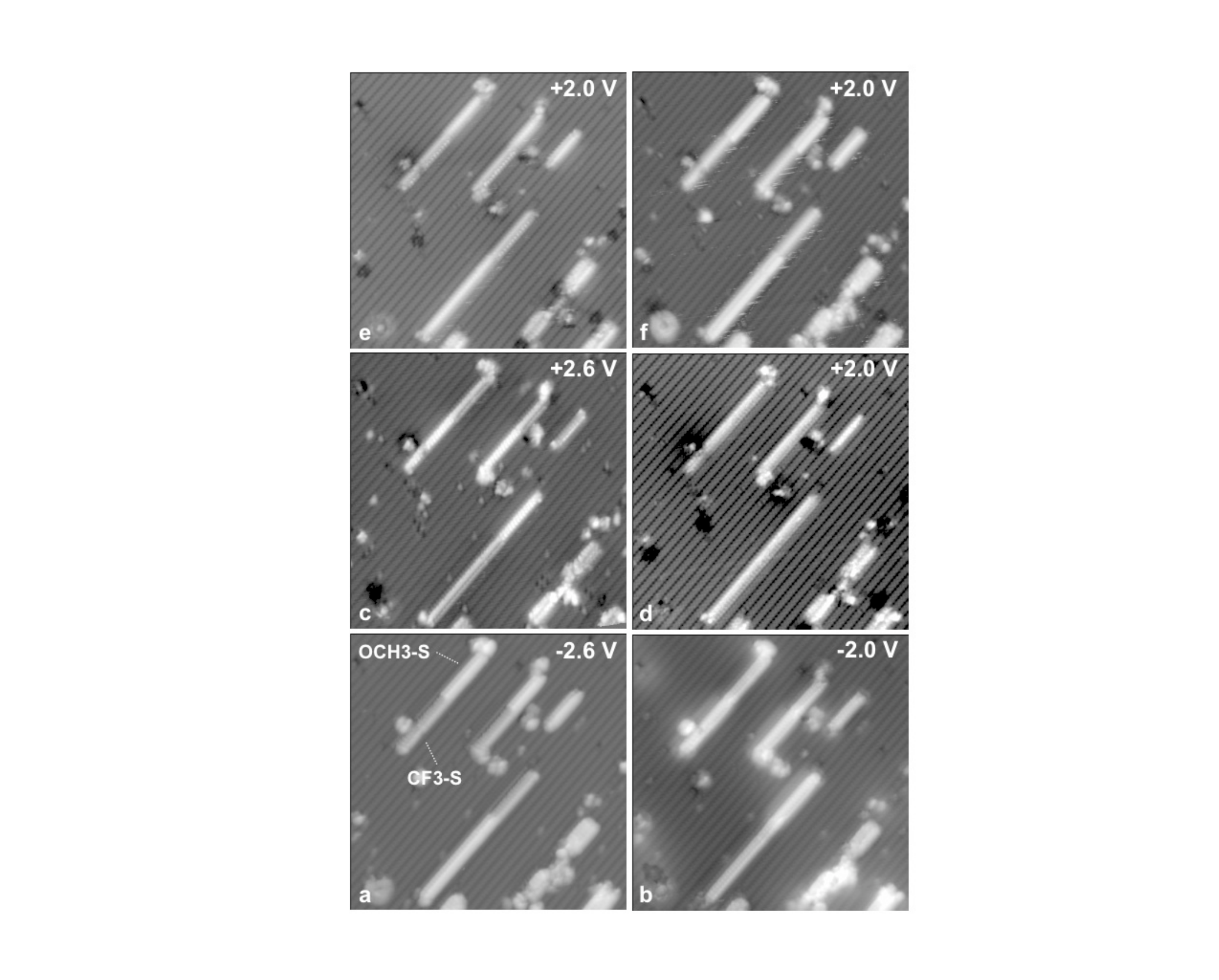}
\caption{\label{Fig_2expt}Constant current filled-state (a)-(b) and empty-state 
(c)-(f) STM imaging of a 3 heterowire cluster. 
(a) $V_s = -2.6$V.  OCH$_3$--styrene line segments image higher (brighter) 
than CF$_3$--styrene line segments. 
(b) $V_s = -2.0$V.  At low magnitude bias OCH$_3$--styrene can image lower 
than CF$_3$--styrene (tip
dependent).  Enhanced molecular conductance throughout the interfacial 
OCH$_3$--styrene remains evident.  (c)
$V_s = +2.6$V.  Consistent with the greater electron affinity of 
CF$_3$--styrene, CF$_3$--styrene images with
increased height relative to OCH$_3$--styrene.  (d) $V_s = +2.0$V.  At 
reduced bias, height contrast between the
CF$_3$--styrene and OCH$_3$--styrene line segments decreases, and 
molecules image with similar corrugation.  (e)
and (f) $V_s = +2.0$V.  Depending on tip structure, OCH$_3$--styrene can 
image above CF$_3$--styrene and with
varying (tip dependent) corrugation.  Greyscale: 0 nm (black), and 0.59nm, 
0.44nm, 0.28nm, 0.18nm, 
0.39nm, 0.18nm (white), (a) to (f), respectively.  Image areas (a)-(f): 26nm$
\times$26nm.  Tunnel current:
40pA}
\end{figure*}

\begin{figure*}[b]
\includegraphics[width=0.70\linewidth, clip=true, trim=0.0 160.0 0.0 100.0]{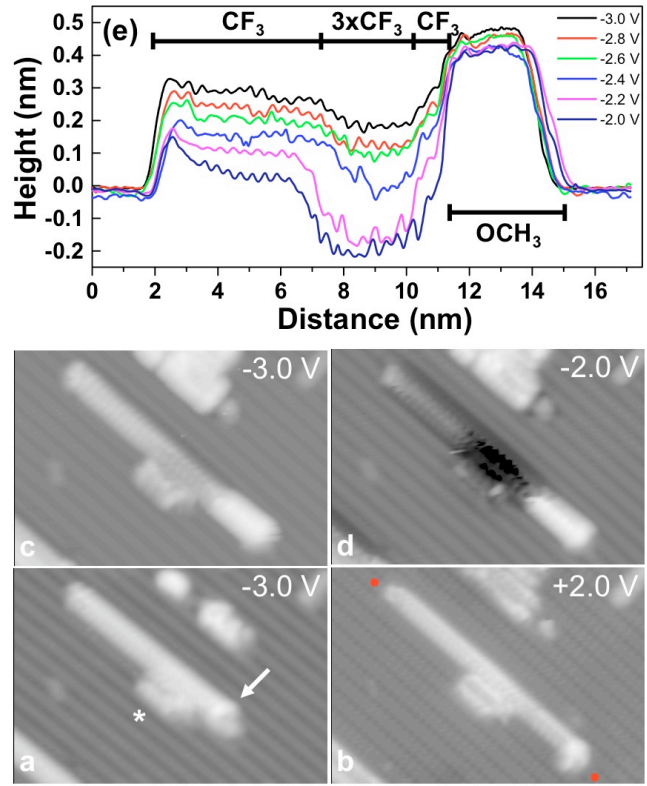}
\caption{\label{Fig_3expt}Color online.  Constant current filled-state STM 
images showing the growth of a (single-triple CF$_3$--styrene)/OCH$_3$--styrene 
heterostructure.  (a) $V_s = -3.0$V.  H:Si(100) surface following a 10 L 
exposure of CF$_3$--styrene.  White arrow indicates the position of the chemically 
reactive dangling bond at the end of the long CF$_3$--styrene segment.  The 
white asterisk shows a short double CF$_3$--styrene line segment beside the 
longer single CF$_3$--styrene chain.  (b) $V_s = +2.0$V. Following 
a 10 L exposure of 
OCH$_3$--styrene, the long CF$_3$--stryene chain has been extended by $\sim 7$ 
OCH$_3$--styrene molecules.  (c) $V_s = -3.0$V. OCH$_3$--styrene images above 
CF$_3$--styrene.  Single and triple CF$_3$--styrene segments image with similar 
height.  (d) $V_s = -2.0$V.  Single chains of OCH$_3$--styrene and CF$_3$--styrene 
continue to image above the H:silicon surface (brighter).  Triple chains of 
CF$_3$--styrene image below the H:silicon surface (black).  (e) Constant 
current topographic cross-sections (0.4nm wide) extracted from bias 
dependent imaging of the 
CF$_3$--styrene/OCH$_3$--styrene heterowire along the trench to the right of 
the attachment dimers.  Heights are given relative to the H:silicon 
surface (height = 0 nm).  Image areas (a)-(d): 15nm$\times$10nm.  
Tunnel current: 40 pA.}
\end{figure*}

H-termination of the Si(100) samples, growth of the molecular 
heterowires and their imaging with an Omicron STM1 
were all carried out within the same vacuum system.
Dissolved atmospheric gases were removed from the 
substituted styrene precursors using
multiple freeze-pump-thaw cycles prior to introduction (via leak valve) into the 
vacuum system.  Before and
after line growth, samples were imaged in STM using electrochemically 
etched tungsten tips, cleaned by
electron bombardment, and field emission.  Bias dependent STM imaging of 
the structures was performed in
constant-current mode (tunnel current fixed at 40 pA).  Surface coordinates 
belonging to STM images in Figures
\ref{Fig_1expt} to \ref{Fig_3expt} were rescaled after acquisition to yield an 
orthogonal inter-row (0.768
nm) to inter-dimer (0.384 nm) periodicity of 2:1.  Feature heights in the 
constant-current STM images were
determined in relation to observed terrace height separation 
on the Si(100) surface 
(0.136 nm).  Images of
CF$_3$--styrene/OCH$_3$--styrene heterowires were acquired on multiple 
surfaces and studied using several
different tips in two separate Omicron STM1 systems.

Figure \ref{Fig_1expt} shows the growth and bias dependent STM imaging of two 
CF$_3$--styrene/OCH$_3$--styrene heterowires on
an H:silicon sample.  After imaging the unreacted H:silicon surface (not 
shown), the STM tip was retracted
${\sim}1$ micron from the surface.  CF$_3$--styrene was introduced into the 
chamber at a pressure of
1$\times$10$^{-6}$ Torr for ${\sim}10$ seconds.  Sample exposure was 
nominally 10L (1L = 10$^{-6}$ Torr
sec). The H-terminated silicon dimer rows run diagonally between 
the upper left and lower right hand corners of the image frames.  
The elongated white (elevated) features running along the dimer rows 
correspond to regions where molecules have reacted with the surface.  
The black (depressed) features appearing most notably in the centers of 
Figs \ref{Fig_1expt} (a) and (b) result from missing atoms (Si vacancies) 
in the silicon surface.

Fig.\ref{Fig_1expt}(a) shows a 26nm$\times$26nm region of the sample 
following the 10L exposure of
CF$_3$--styrene.  Sample bias ($V_s$) was $-3.0$V.  White arrows label the 
reactive dangling bonds at the end of two
CF$_3$--styrene line segments.  Due to slight tip asymmetry, CF$_3$--styrene 
bound to either side of their host dimers
image with slightly different corrugation.  Comparison with images of the 
unreacted H:silicon surface (not
shown) show the upper and lower CF$_3$--styrene line segments are 
chemically bound to the right-hand and left-hand
sides of their respective dimer rows.

Fig.\ref{Fig_1expt}(b) shows the same region of the crystal ($V_s = -3.0$V) 
following a 10L exposure of OCH$_3$--styrene
(1 $\times$ 10$^{-6}$ Torr exposure for ${\sim}10$ seconds).  The terminal 
dangling bonds belonging to the upper and
lower CF$_3$--styrene line segments in (a) have reacted with 
OCH$_3$--styrene forming two CF$_3$--styrene/OCH$_3$--styrene heterowires
(`1' and `2').  At $-3.0$V, the tip Fermi-level is below the highest band of 
occupied molecular $\pi$ states for the OCH$_3$--styrene as evidenced by 
the fact that at this bias voltage  the experimental STM height profile of the 
OCH$_3$--styrene has saturated as can be seen in Fig.1(d): This saturation 
indicates that the number of OCH$_3$--styrene HOMO states
contributing to the STM current is no longer increasing with increasing bias so
that the STM tip Fermi level must be below the highest band of 
OCH$_3$--styrene HOMO levels.
However, the tip Fermi-level remains above the occupied molecular $\pi$ 
states in the CF$_3$--styrene at this bias because the molecular 
CF$_3$--styrene HOMO is well below the OCH$_3$--styrene 
HOMO as evidenced by gas 
phase molecular calculations\cite{DiLabio99} and by the present theoretical 
work for these molecules on silicon. This is also consistent with the CF$_3$-
styrene
molecules imaging with reduced height (less bright) in comparison to the 
OCH$_3$--styrene in Fig.\ref{Fig_1expt}.

Fig.\ref{Fig_1expt}(c) shows the same region of the crystal imaged with $V_s$ 
= -1.8 V.  At this bias, the tip
Fermi-level is in the vicinity the highest occupied molecular orbitals ($\pi$ 
states) in the OCH$_3$--styrene line segment; this conclusion is based on 
comparisons between these experimental data and the results of our theoretical 
modeling as is discussed in Section \ref{Positive}.      
While the OCH$_3$--styrene continues to image above (brighter than) the 
CF$_3$--styrene, heterowires 1 and 2 image with enhanced
height (brighter) above the OCH$_3$--styrene molecules situated near the 
heterointerface.  The OCH$_3$--styrene molecules in
heterowire 1 (close-up in inset) near the terminal dangling bond also image 
with increased height. 

Fig.\ref{Fig_1expt}(d) shows a series of 0.4nm wide topographic 
cross-sections extracted from heterowire 1 along the
trench running between its attachment dimer row (labelled with red circles in 
Fig.\ref{Fig_1expt}(c) inset) and the
vacant H-terminated dimer row to its right. 
Also included are curves from images (not 
shown) acquired at intermediate sample
biases.  The topographic height envelope for the heterostructure extends 
between 1 nm and 9.5 nm along
the abscissa.  The height maxima associated with the terminal dangling bond, 
and the molecular heterojunction are
located at
2.3 nm, and 6.4 nm, respectively.  Postponing discussion of the 
interfacial height enhancement for the time
being, the bias dependent height response of the OCH$_3$--styrene line 
segment near the terminal dangling bond is much like
that reported in Ref.\onlinecite{Nature} for a single homomolecular styrene 
line segment:  At elevated negative substrate bias, ($|V_s|
\gtrsim 2.4$ V) the highest band of occupied $\pi$ states in the 
OCH$_3$--styrene lies above the tip Fermi-level and the
OCH$_3$--styrene images with roughly uniform height.  As the bias decreases 
in magnitude, these occupied $\pi$ states of
the OCH$_3$--styrene approach the tip Fermi-level (and eventually begin to 
drop below it), and the molecules image with
reduced height above the surface.  Molecules in the vicinity of the negatively 
charged dangling bond on the degenerately
doped n-type surface exhibit a spatially dependent reduction in ionisation 
potential due to the electrostatic field
emanating from the negatively charged dangling bond.\cite{Nature,Raza}  
Thus at low bias, molecules nearest the terminal
dangling bond image with increasing height.\cite{Nature}

In stark departure from the imaging characteristics of homomolecular styrene 
chains reported in
Ref.\onlinecite{Nature}, and absent from the images of the 
CH$_3$--styrene/styrene heterowires 
studied in Ref.\onlinecite{stymeth} is the
height enhancement at the heterojunction resolved at low bias.  
The bias-dependent height response of the OCH$_3$--styrene near the
heterojunction is similar to that near the terminal dangling bond just 
described.  At elevated bias, the interfacial OCH$_3$--styrene images with 
nearly constant height along the bulk
of the homowire segment.  As $|V_s|$ decreases, however, the height of the 
interfacial OCH$_3$--styrene (4-5 molecules
closest to the molecular interface) does not decay as rapidly as in the 
mid-section of the OCH$_3$--styrene line segment. 
At the lowest filled-state bias studied for this particular heterowire, the 
interfacial OCH$_3$--styrene molecules image
${\sim}0.05$ nm higher than OCH$_3$--styrene molecules situated 5-7 dimers 
away from the heterojunction.
The typical noise in the presented images is $<0.004$ nm.

Fig.\ref{Fig_2expt} shows filled and empty-state imaging of a cluster of 3 
heterowires.  These heterowires were
identified on a region of the crystal where imaging of the H:silicon surface 
before line growth was not carried out.  The
chemical identities of the line segments were confirmed at the end of the 
imaging sequence by dosing with CF$_3$--styrene, and
comparing the imaging characteristics of the newly reacted line segments (not 
shown) with the nearby heterowires. 
Characteristics in these images (in combination with those in Fig.\ref
{Fig_1expt}) are representative of the
range of tip-dependent imaging contrast encountered 
over the course of the experiments.

Fig.\ref{Fig_2expt}(a) shows a filled-state constant current STM image of the 3 
heterowire cluster with $V_s = -2.6$V.  As
in Fig.\ref{Fig_1expt}(b), the OCH$_3$--styrene line segments image taller 
(brighter) than the CF$_3$--styrene line segments
consistent with their relative ionisation potentials.  In Fig.\ref{Fig_2expt}(b) 
($V_s = -2.0$V), however, the
OCH$_3$--styrene away from the interface images lower (darker) than the 
CF$_3$--styrene.  This contrast reversal (tip
dependent) is often encountered at low bias in filled-state imaging.  In all 
instances, whether or not this low bias
contrast reversal occurs, height enhancement of the interfacial 
OCH$_3$--styrene remains prominent and corresponds to the height
of OCH$_3$--styrene far from the interface at greater magnitude filled-state 
bias (i.e. the interfacial OCH$_3$--styrene
images as though accompanied by a localised increase in effective 
tip-sample bias magnitude). 

Figs \ref{Fig_2expt}(c) to (f) show empty-state images for the 3 heterowire 
cluster.  In Fig.\ref{Fig_2expt}(c) at $V_s
= +2.6$V, the CF$_3$--styrene images above (brighter than) the 
OCH$_3$--styrene.  As is discussed
in Section \ref{Positive} this result is consistent with the
increased electron affinity of the CF$_3$--styrene (a result of the highly 
electronegative F atoms).  As the $\pi$* state
for the CF$_3$--styrene is much lower than the corresponding $\pi$* states 
belonging to the OCH$_3$--styrene, a greater
number of tip states can tunnel into the CF$_3$--styrene (and therefore, the 
STM tip must travel further away from the surface
above the CF$_3$--styrene to maintain a constant tunnel current across the 
heterostructure).  In Fig.\ref{Fig_2expt}(d),
$V_s = +2.0$V and the tip Fermi-level is presumed to lie below both the 
OCH$_3$--styrene (as before) and CF$_3$--styrene
$\pi$* states.  The tunnel current is dominated by carrier injection into the 
silicon conduction band, and the molecules
image (non-resonantly) with similar contrast.

Figs \ref{Fig_2expt}(e) and (f) display empty-state imaging data acquired at 
$V_s$ = +2.0 V following spontaneous tip structural changes.  In
this imaging mode, the OCH$_3$--styrene images above (brighter than) the 
CF$_3$--styrene.  This imaging behaviour when apparent,
often extends to greater bias (up to $V_s$ = +3.2 V).  In addition to this 
empty-state contrast reversal, changes to the
molecular corrugation within the homomolecular line segments are apparent 
(particularly in terms of the appearance of the
OCH$_3$--styrene between frames (d) and (f)).  It is clear in these experiments 
that the tip density of states plays a
considerable role in determining the contrast (in terms of overall height and/or 
molecular corrugation) observed between
the OCH$_3$--styrene and CF$_3$--styrene line segments.  Tip-dependent 
imaging contrast in STM of organic molecules has
been documented previously.\cite{Cyr,Hahn}

While the empty-state imaging displays considerable contrast variation, of 
particular significance is the absence of
notable interfacial structure.  An exception to this was observed in a small 
portion of low bias empty-state images
which revealed height enhancement for a single interfacial molecule.  The 
height enhancement was between 0.01 to 0.02
nm, and smaller than the 0.02 nm to 0.14 nm height enhancement observed 
for the considerably broader ($\sim 5$ molecule wide)
filled-state interfacial features resolved in Figs \ref{Fig_1expt}(c)-(d), \ref
{Fig_2expt}(b), and others (not shown). 
While such a narrow empty-state feature appears in the simulation work 
presented in Sections \ref{Case} and
\ref{Relationship}, additional observations will be required before a 
detailed comparison with theory can be undertaken.

Fig. \ref{Fig_3expt} shows STM imaging results for a 
CF$_3$--styrene/OCH$_3$--styrene heterostructure with multiple 
CF$_3$--styrene chains in a side by side configuration.  
Fig. \ref{Fig_3expt}(a) shows a region of the H:silicon surface following a 
10 L exposure of CF$_3$--styrene ($V_s = -3.0$ V). The arrow points to the 
reactive dangling bond at the end of the longest CF$_3$--styrene line. 
The $\star$ marks a short double chain of CF$_3$--styrene that has grown 
beside the long CF$_3$--styrene chain. Figures \ref{Fig_3expt}(b) to 
\ref{Fig_3expt}(d) show the same region following a 10 L exposure of 
OCH$_3$--styrene. The end of the long CF$_3$--styrene chain has been 
extended by approximately 7 molecules of OCH$_3$--styrene. Figures 
\ref{Fig_3expt}(b) ($V_s = +2.0$V) and \ref{Fig_3expt}(c) ($V_s = -3.0$V) 
image the single and triple CF$_3$--styrene segments with comparable 
height. In Fig. \ref{Fig_3expt}(d), $V_s$ has been reduced to $-2.0$ V and 
the region with the triple CF$_3$--styrene lines images below (darker than) 
the single file chain of CF$_3$--styrene. Fig. \ref{Fig_3expt}(e) shows 
topographic cross sections along the CF$_3$--styrene/OCH$_3$--styrene 
heterowire. From $V_s = -3$ V to $V_s = -2$ V, the triple CF$_3$--styrene 
chain 
(between 7 and 10 nm along the abscissa) images with decreasing height. 
Significantly, at $V_s =-2.0$ V, this region images 0.2 nm below the H:Si 
surface indicating depleted silicon state density beneath the molecules 
at the tip Fermi level.

Results presented in Figures \ref{Fig_1expt} and \ref{Fig_2expt} and for 
other single chain CF$_3$--styrene/OCH$_3$--styrene heterostructures 
(not shown), show the presence of the chemical heterojunction causes the 
interfacial OCH$_3$--styrene to image with elevated height under low 
filled-state bias.  The prominence (both in terms of height and lateral 
extent) of the filled-state interfacial feature compared with the absence 
of significant interfacial structure resolved in empty-state imaging in these 
structures, suggests an electronic origin for the effect.  As the low bias 
imaging conditions required to observe the filled-state interfacial feature 
also result in decreased tip-sample separation, the possibility of tip-molecule 
interactions at the heterojunction leading to altered molecular conformations 
which contribute to the observed interfacial structure cannot be ruled out. 
However, such tip-induced structural changes cannot account 
for the differing interfacial behaviour observed at low 
positive and negative substrate bias.
The localized depletion in silicon filled-state density observed in 
Fig. \ref{Fig_3expt}  (typical of other side by side CF$_3$--styrene 
structures studied) in response to the triple CF$_3$--styrene chains 
also cannot be understood in terms of only conformational 
differences between 
otherwise non-interacting molecules.  
The theoretical modeling developed 
in the following Sections explores various factors which can account for 
these observations. It will be seen that conformational details 
in concert with electrostatics play a significant role in these phenomena.
\section{Model}
\label{Model}

\subsection{Formalism}
\label{Formalism}
 
In order to carry out calculations of electronic transport through
molecules bonded chemically to metal or semiconductor electrodes it is 
necessary to
know the electronic structures of these systems. {\it Ab initio} density
functional calculations based on the Kohn-Sham local density approximation 
(LDA) \cite{KohnSham}
and extensions such as generalized gradient approximations (GGA) are 
commonly used for this purpose. However, the theoretical foundations of this 
approach and the accuracy of the results that it yields for molecular {\em 
transport} calculations are increasingly being questioned in the literature at 
the present time:\cite
{Delaney,Sai,Toher05,Darancet,Evers,Ke07,Pemmaraju,Prodan,offsets,
Koentoppreview,Thygesen78,Toher78,Thygesen8,GK} 
While such formalisms are appropriate (at least in principle) for 
calculating the {\it total} energy, the
spatial distribution of the electronic charge density and the electrostatic 
potential throughout
inhomogeneous  electronic systems in their {\em ground states},\cite
{KohnSham, 
HK} the single-electron eigenenergies and wave functions that appear in 
them are somewhat artificial constructs that 
in most cases do not have rigorous physical meanings.\cite
{KohnSham,GK,Evers,Koentoppreview} Consequently, they need not be good 
approximations to the energies and wave functions of the true electronic 
quasi-particles that
determine the electronic transport properties of molecular systems. 
Under what circumstances such {\it ab
initio} calculations should produce acceptable results for electron transport 
{\em despite} this
fundamental shortcoming, and how to obtain satisfactory results when they do 
not are important
unresolved questions that are the subject of much ongoing research at the 
present time.\cite
{Delaney,Sai,Toher05,Darancet,Evers,Ke07,Pemmaraju,Prodan,offsets,
Koentoppreview,Thygesen78,Toher78,Thygesen8,
GK,Buker}

For molecules adsorbed on silicon the above deficiencies of the LDA, GGA 
and their relatives
manifest themselves most obviously in that these approximations 
underestimate the band gap of silicon
and yield incorrect values for the energy offset between the highest occupied 
molecular orbital
(HOMO) of the molecule (or the relevant frontier orbital of the adsorbate) and 
the silicon valence band
edge. The errors in these energy offsets obtained from the density functional 
calculations have
recently been estimated for a few molecules to range from 0.6 to 1.4eV.\cite
{offsets} Because
of these and other\cite{Rakshit} deficiencies, the predictions of transport 
calculations based on {\it ab initio} density
functional calculations of the electronic structure are unreliable for molecules 
on
silicon; they are able to capture some observable phenomena\cite{Bevan} but 
are qualitatively
incorrect for others.\cite{offsets,stymeth}

In the case of linear chains of styrene\cite{Lopinski} and methyl-styrene\cite
{stymeth} molecules
grown by self-assembly on a hydrogenated (001) silicon surface, the incorrect 
offsets between
the molecular HOMO levels and the silicon valence band edge given by the 
density
functional calculations result in such calculations yielding qualitatively 
incorrect STM images
for these systems.\cite{stymeth} In particular, at low bias the {\it ab initio} 
calculations predict
{\em minima} in the STM height profiles of the molecular chains over the 
centers of the molecules
where {\em maxima} are observed experimentally.\cite{stymeth} These 
deficiencies of the
{\it ab initio} calculations have been overcome\cite{stymeth} by developing a 
different electronic
structure model based on extended H{\"u}ckel theory, a tight
binding scheme from quantum chemistry \cite{Hoffman,Ammeter} that 
provides an approximate description of the electronic
structures of many molecules and has also been used successfully to explain 
the experimental
current-voltage characteristics of a variety molecular
wires connecting metal electrodes\cite
{Datta,Emberly,EmberlyPRB,Kushmeric,Cardamone} and
to model the band structures of a variety of crystalline solids \cite
{stymeth,Cerda,Kienle}
 
Extended H{\"u}ckel theory describes molecular systems in terms
of a small set of Slater-type atomic orbitals $\{
|\phi_i\rangle \}$, their overlaps $S_{ij} =
\langle\phi_i | \phi_j\rangle$ and a Hamiltonian matrix
$H_{ij} =
\langle\phi_i |H| \phi_j\rangle$. The diagonal
Hamiltonian elements $H_{ii} = \epsilon_i$  are taken to be 
the atomic orbital ionization energies and the non-diagonal
elements $H_{ij} $ for $i \ne j$ are expressed in terms of  $\epsilon_i$, $
\epsilon_j$,  $S_{ij}$
and phenomenological parameters chosen
for consistency with experimental molecular
electronic structure data. 
As is described in detail in Ref. \onlinecite{stymeth},
the standard extended H{\"u}ckel theory\cite{Ammeter}  was modified so as to 
also provide an accurate description of
the band structures of the silicon substrate and the tungsten STM tip.  The 
energy offset between the molecular HOMO and the silicon 
valence
band edge (an adjustable parameter in the theory\cite{stymeth}) was 
assigned a physically reasonable value for which our transport calculations 
reproduced correctly the character of the height profile
along the styrene and methylstyrene molecular chains observed 
experimentally in STM images at low bias,
i.e., apparent height {\em maxima} over the centers of the molecules. The 
model proved to be remarkably
successful, accounting not only for this low bias property of the STM images, 
but also for several 
counter-intuitive features of the experimental data, including the 
experimentally observed reversal in
the contrast between the styrene and methyl-styrene molecular chains with 
increasing STM tip bias, the
observed increase in the apparent height of the molecules at the ends of the 
molecular chains relative
to those far from the ends with increasing bias and the observed 
disappearance of the corrugation of
the STM height profile along the molecular chains with increasing bias.\cite
{stymeth} 

The tight binding model based on extended H{\"u}ckel theory that is described 
in Ref.
\onlinecite{stymeth} is adopted in the present work but with an important 
modification: 
A limitation of extended H{\"u}ckel theory is that in it the atomic orbital energy
$\epsilon_i$ depends on the type of atom on which orbital $i$ is located but is
not influenced by the presence of other atoms in the vicinity. This is a 
reasonable first approximation
for the styrene and methylstyrene molecules considered in Ref. \onlinecite
{stymeth} since those
molecules do not contain strongly charged groups. However for the OCH$_3
$-styrene and CF$_3$--styrene
molecules that will be considered here, there is strong charge transfer 
between carbon atoms and
fluorine and oxygen atoms that results in significant electrostatic fields that 
should
modify the atomic orbital energies
$\epsilon_i$ of surrounding atoms. These electrostatic fields {\em are
included} in the present model as is explained below. The resulting variation 
in the electrostatic potential from molecule
to {\em like} molecule along the molecular chains has direct and striking 
effects on the experimental STM height profiles presented in Section \ref
{Expt}, as will be discussed in detail 
in the Sections that follow.

While, as has been discussed above and in Refs. 
\onlinecite
{Delaney,Sai,Toher05,Darancet,Evers,Ke07,Pemmaraju,Prodan,offsets,
Koentoppreview,Thygesen78,Toher78,Thygesen8,
GK}, the use of density functional theory at the level of LDA
or GGA for calculating electronic quasiparticle properties has no fundamental 
justification, this
criticism does not apply to calculations of the ground state electronic charge 
density
distribution {\em and the electrostatic potentials} that that have been shown by 
Hohenberg and Kohn\cite{HK} to be functionals of the charge density.
Therefore the use of  {\it ab initio} density functional calculations to estimate 
the electrostatic contributions to the
atomic orbital energies
$\epsilon_i$ that are due to charge transfer between different atoms (but are 
neglected in extended
H{\"u}ckel theory), while still involving approximations, is justified at the 
fundamental level. 
In the present work these electrostatic
corrections to extended H{\"u}ckel theory were included in our model in the 
following way: 

An 
{\it ab initio} calculation was carried out\cite{Gaussian} of the electrostatic 
potential $W_n$ at the
nucleus of every atom $n$ of an atomic cluster that included a molecular 
chain with a total of 20
OCH$_3$--styrene and CF$_3$--styrene molecules in the geometry that they 
take on the silicon surface, a few
layers of nearby Si atoms, and the H atoms needed to passivate the dangling 
bonds on the surface of
this silicon cluster. A similar {\it ab initio} calculation was carried out\cite
{Gaussian}
of the electrostatic potential $U_n$ at the nucleus of each of these atoms in 
the {\em absence} of all
of the other atoms. Thus 
\begin{equation}
E_n = -e(W_n - U_n) 
\label{electrostics}
\end{equation}
is an estimate of the contribution to the electrostatic
energy of an electron in an atomic orbital on atom $n$ that is due to the 
presence of all of the
{\em other} other atoms in the system, calculated self-consistently from first 
principles. I.e.,
it is the electrostatic contribution to the atomic orbital energy that is neglected 
in extended
H{\"u}ckel theory, as discussed above. This contribution was included in the 
present model by
making the substitution $\epsilon_i \rightarrow \epsilon_i + E_n$ (for each 
atomic orbital $i$ of every
atom $n$) in the diagonal elements $H_{ii} =
\epsilon_i$ of the extended H{\"u}ckel-like model Hamiltonian obtained for this 
system as described in
Ref.\onlinecite{stymeth}. Because the orbital basis used is not orthogonal (as 
in standard extended
H{\"u}ckel theory), the non-diagonal matrix elements of the model Hamiltonian 
were also adjusted
according to
$H_{ij} \rightarrow H_{ij} + S_{ij}(E_n + E_m)/2$ for orbitals $i$ and $j$ on 
atoms $n$ and $m$ as
required for gauge invariance.\cite{shift} The effect of the bias voltage applied 
experimentally
between the STM tip and the silicon substrate on the Hamiltonian matrix $H_
{ij}$ was included in the
present model as in Ref.\onlinecite{stymeth}. 

Since in the above theoretical approach a highly compute-intensive {\em ab 
initio}
calculation of the effects of charge transfer is carried out only once for any 
molecular chain
being studied and a smaller basis set is used in the transport calculations that 
follow, in
addition to correcting some fundamental deficiencies of density functional 
theory as
discussed above, the present approach is able to treat much larger systems 
than is practical to
study by {\em ab initio} methods alone. This was crucial for the present work 
where fairly
long molecular chains on Si needed to be studied and the system is not 
periodic along the molecular chain.

The electric current flowing between the
STM tip and silicon substrate via the adsorbed molecules was evaluated as 
described in detail in Ref.
\onlinecite{stymeth} by solving the Lippmann-Schwinger equation, 
determining from the solution the
Landauer transmission probability $T(E,V)$ that depends on the electron 
energy $E$ and applied bias
voltage $V$ and evaluating the Landauer formula    
\begin{equation}
I(V) = \frac{2e}{h} \int_{-\infty}^{+\infty}
dE\:T(E,V)\left( f(E,\mu_{s}) - f(E,\mu_{d})\right)
\label{Landauer}
\end{equation}
where $I$ is the current, $f(E,\mu_{i}) ={1}/{(\exp[(E-\mu_{i})/kT] + 1)}$
and $\mu_{i}$ is the electrochemical potential of the
source ($i=s$) or drain ($i=d$) electrode.\cite{temp}

The chains of OCH$_3$--styrene and CF$_3$--styrene molecules and 
associated silicon clusters that needed to be 
studied theoretically to make comparison with the experiment included too 
many atoms
for it to be practical to determine their relaxed structures using $ab$ $initio$ 
density
functional theory calculations. Therefore a molecular mechanics method was 
adopted: Chains
of 40 molecules, half of the chain OCH$_3$--styrene and the other half 
CF$_3$--styrene, on a hydrogenated
silicon cluster were relaxed using the Universal Force Field model\cite
{GaussianUFF} starting from a
variety of initial configurations, and many different metastable relaxed 
structures were found.
It is reasonable to
suppose that many of them as well as intermediate structures between 
them were being sampled thermally in the
experiment which was carried out at room temperature. Thus the limited 
accuracy of the structures given
by the molecular mechanics method was deemed to be adequate 
for the purpose 
of the present study. Since the
phenomena of interest in the present work were observed experimentally 
near the junction of the
OCH$_3$--styrene and CF$_3$--styrene chains, the relaxed molecular chain 
was truncated to a subset of 20
molecules surrounding the junction and the calculations of the electrostatic 
potential and of
electrical conduction between the  silicon substrate and STM tip via the 
molecules were carried out for
this truncated chain of molecules and underlying H-terminated Si cluster with 
a dimerized (001) surface without further relaxation. The STM tip in the 
present work was modelled as in
Ref.\onlinecite{stymeth} as a 15 tungsten atom cluster with a (001) orientation 
and a single
terminating atom, and coupled to an electron reservoir by many ideal leads.  

\subsection{Energy Level Ordering}
\label{Level Ordering}

In the present
theoretical approach, the energy offsets between the HOMO levels of the 
OCH$_3$--styrene and CF$_3$--styrene 
parts of the molecular chain are given by 
the extended H{\"u}ckel theory with the  parameterization described in Ref. 
\onlinecite{Ammeter}, modified so as to yield an accurate band structure for 
silicon and to  include the {\em ab initio} electrostatic corrections $E_n$ as 
discussed in Section \ref{Formalism}. The same applies to the molecular  
LUMO levels and
to the variation of the atomic orbital energies from molecule to molecule along 
the molecular chain.

The molecular HOMO and LUMO levels for both types of molecules 
considered
here reside primarily on the benzene rings of the molecules. Thus,
because of the electron withdrawing (donating) nature of the 
CF$_3$ (OCH$_3$)
group, the OCH$_3$--styrene HOMO is higher in energy than the 
CF$_3$--styrene HOMO,\cite{DOS}
and the CF$_3$--styrene LUMO is lower than the OCH$_3$--styrene LUMO.
Thus the HOMO of the molecular chain as a whole is located on its 
OCH$_3$--styrene
part (as has already been mentioned in Section \ref{Expt}) and the LUMO of 
the molecular chain as a whole is located on its CF$_3$--styrene
part.

However, as in Ref. \onlinecite{stymeth}, the energy offset between the 
molecular HOMO of the molecular chain and the silicon valence band edge is 
an adjustable parameter of the present theory whose value is not known 
accurately: As was discussed at the beginning of Section \ref{Formalism}, this 
offset is not given correctly by calculations based on standard density 
functional theories; its accurate
determination (like the determination of the energy offsets between the 
molecular levels and substrate Fermi levels for other molecular nanowires\cite
{Ratner06,GK}) is a difficult unsolved problem of molecular electronics. 
However,
based on the discussion of our experimental results that follows it is plausible 
that the molecular HOMO is located below the silicon valence band edge and 
the molecular LUMO is
located above the silicon conduction band edge:  

An important aspect of the experimental data presented in Section \ref{Expt} 
is that even for the smallest values
of $|V_s|$, the bias voltage voltage between the STM tip and substrate at
which imaging of the molecular chains was feasible, $|eV_s|$
was considerably larger than the band gap of silicon, for {\em both} positive
and negative biasing of the silicon substrate relative to the STM tip.
Given that the Fermi level of the STM tip at zero bias is located within the Si 
band gap, it therefore follows that 
the STM tip Fermi level was well below the top of the silicon valence 
band or well above the bottom of the silicon conduction band when
the molecules were being imaged for negative and positive biasing of
the substrate relative to the STM tip, respectively. 
A reasonable interpretation 
of this 
fact (that is consistent with all of the
experimental data and with the results of our theoretical modeling)
is that the HOMO of the molecular
chain is located in energy below the silicon valence band edge,
that the LUMO of the molecular
chain is above the silicon conduction band edge, and that the
experimental imaging
of the molecular chains themselves (in the present study)
was being carried out for STM tip Fermi energies
near or above the LUMO or near or below the HOMO of the molecular chain, 
conditions
under which enhancement of  STM 
currents due to resonant or near resonant transport via the molecules is to be 
expected. This will
be the view adopted in the remainder of this article. 

Thus three regimes will be considered theoretically
for both positive and negative substrate bias: 

1. Very low bias at which the STM tip Fermi level lies between the
Si conduction band minimum (valence band maximum) and the molecular 
LUMO (HOMO) energy levels.

2. Low bias at which the STM tip Fermi level is slightly above (below)
the lowest (highest) molecular LUMO (HOMO) level.

3. Higher bias at which the STM tip Fermi level is near the top (bottom)
of the lowest (highest) molecular LUMO (HOMO) band.

As will be seen below, each regime is predicted to have its own characteristic 
signature
in STM imaging that can be compared with our experimental data. These 
signatures do not depend qualitatively on the precise value assumed for the 
offset between the silicon valence
band edge and molecular HOMO level. Thus comparison between theory and 
experiment
allows us to draw conclusions regarding the specific regimes in which the 
data was being taken
and the physical mechanisms underlying the phenomena that were observed.

\subsection{Prototypical Geometries of the Individual Molecules Bound to Si 
Dimers}
\label{Geom}

\begin{figure*}[b]
\includegraphics[width=1.00\linewidth, clip=true, trim=0.0 120.0 0.0 0.0]{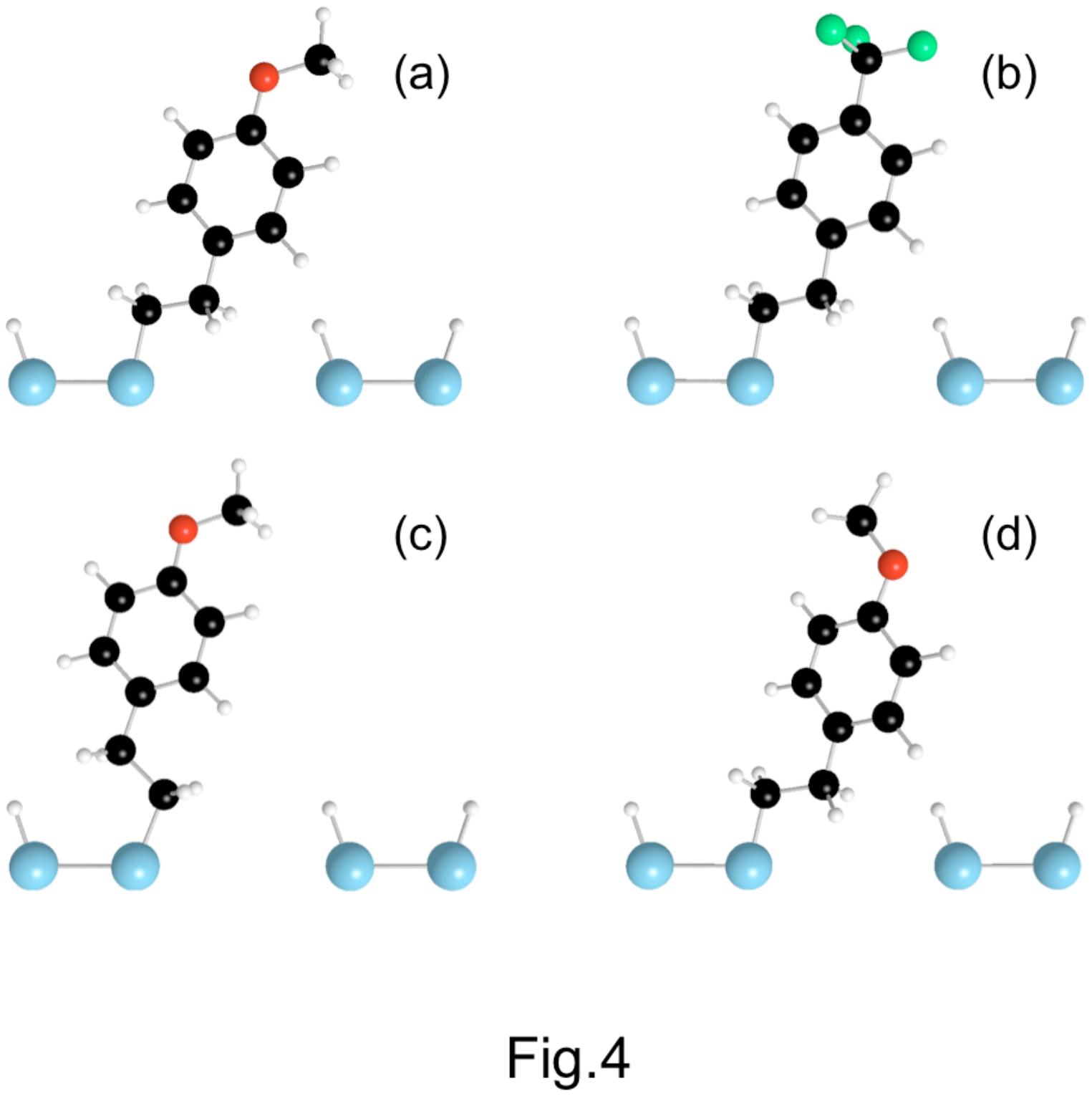}
\caption{\label{Fig_4struct}Color online. Schematic representation\cite{Macmolplt}
of some prototypical geometries of OCH$_3$--styrene (a,c,d) and CF$_3$-
styrene (b)  molecules
on H terminated Si(100). Si,C,O,F and H atoms are blue, black, red, green and 
white
respectively. In each case the molecule is shown together with the two Si 
atoms of the
Si dimer to which the molecule bonds and two Si atoms of an adjacent
Si dimer row. In the T-tethered geometries (a,b and d) the molecule is located 
mainly
over the trench between Si dimer rows while in the D-tethered geometry (c) 
the molecule is mainly
over the Si dimer row to which it is bound. The OC bonds of the OCH$_3$ 
groups are T-oriented (towards the trench) in
(a) and (c) or D-oriented (towards the dimer to which the molecule binds) in
(d). }
\end{figure*}

Different conformations have been proposed for styrene 
molecules  on H terminated Si(100)\cite{Lopinski,stymeth,Cho}           
and it is reasonable to suppose that different atomic geometries
are possible for the substituted styrenes on Si that are studied here. One 
possibility is
that the C atom of the molecule that bonds to the C atom that bonds to the Si is 
located over the
``trench" between two Si dimer rows as is shown in Fig.\ref{Fig_4struct}(a) for  
OCH$_3$--styrene
and in Fig.\ref{Fig_4struct}(b) for CF$_3$--styrene. This conformation is
similar to that assumed for chains of styrene molecules on Si in Refs 
\onlinecite{Lopinski}           
and \onlinecite{stymeth} and will be referred to as ``the T-tethered geometry". 
In another conformation that has also been proposed for styrene on Si\cite
{Cho}
the C atom that bonds to 
the C atom that bonds to the Si is located over the
the Si dimer row to which the molecule bonds as in Fig.\ref{Fig_4struct}(c); 
this alternate conformation will be referred to as ``the D-tethered geometry." 
In Fig.\ref{Fig_4struct}(a) and (c) the C atoms of the OCH$_3$ groups
are located further over the ``trench" than the O atoms are, i.e., the OCH$_3$ 
group
also has the T orientation. However, the opposite (D) orientation of the  OCH
$_3$ group
with the C oriented towards the Si dimer to which the molecule bonds as 
shown in 
Fig.\ref{Fig_4struct}(d) is also possible. 

The molecular conformations described above are
ideal cases: Molecule-molecule interactions in the molecular chains 
and thermal motion at room temperature
are expected to result in many intermediate geometries with the molecular 
benzene ring and
the CO bond of the OCH$_3$ group not being coplanar with each other or
with the silicon dimer to which the molecule bonds and the OCH$_3$ and 
CF$_3$ groups being
rotated through different angles about the bonds that link them to their 
respective molecules.
As will be seen below, these deviations from 
the ideal geometries need to be taken into consideration when modeling the 
experimental
STM images of the molecular heterowires.  However it will often be 
convenient
in the discussions that follow to classify molecular structures  according to the 
idealized structures such as those in Fig.\ref{Fig_4struct} 
(T or D tethered molecule, T or D oriented OCH$_3$ group) that they most 
closely resemble.

\section{Theoretical Results for a representative molecular chain geometry}
\label{Case}

\subsection{Structure}
\label{Structure}

In this Section we present theoretical results for electrical  conduction between 
the STM tip and Si substrate 
through a particular 20-molecule CF$_3$--styrene/OCH$_3$--styrene chain on 
the dimerized H-terminated (001) silicon surface. 
The relaxed geometry of a {\em part} of this molecular chain surrounding the  
CF$_3$--styrene/OCH$_3$--styrene junction is shown at the 
top of Fig. \ref{Fig_5th}. Atoms are colored as in Fig.\ref{Fig_4struct}. 
Each molecule bonds to a Si atom on the same edge of the same Si dimer 
row 
(the far edge of the dimer row closest to the viewer)
through a single C-Si bond as in Fig.\ref{Fig_4struct}. For the specific 
geometry shown in Fig. \ref{Fig_5th}
the tethering of all of the molecules to the Si as well as the orientations of the 
OCH$_3$ groups
are of the T type defined in Section \ref{Geom}. Some further
details of this structure should also be noted since they shall prove to be
crucial in making the connection between the theoretical results that 
will follow and key aspects
of our experimental 
data:  {\em Most}
of the molecules lean somewhat in the direction away from the heterojunction 
and also have swung outwards somewhat (away from the heterojunction) 
about the axes 
of the tethering Si-C bonds, as one might expect for molecules subject
to a net weak mutual steric repulsion.
Due to the repulsion between the
F atoms on different CF$_3$--styrene molecules, the CF$_3$ groups are all 
rotated through
different angles about the bonds between the CF$_3$ groups and the 
molecules'
benzene rings, and the benzene rings themselves are tilted somewhat 
differently from
molecule to molecule. Because of this the detailed structure of 
CF$_3$--styrene
chain is more complex and less well ordered 
than that of the OCH$_3$--styrene chain.

\subsection{Theoretical current profiles at positive and negative substrate bias}
\label{Positive}
\begin{figure*}[b]
\includegraphics[width=0.90\linewidth, clip=true, trim=0.0 150.0 0.0 100.0]{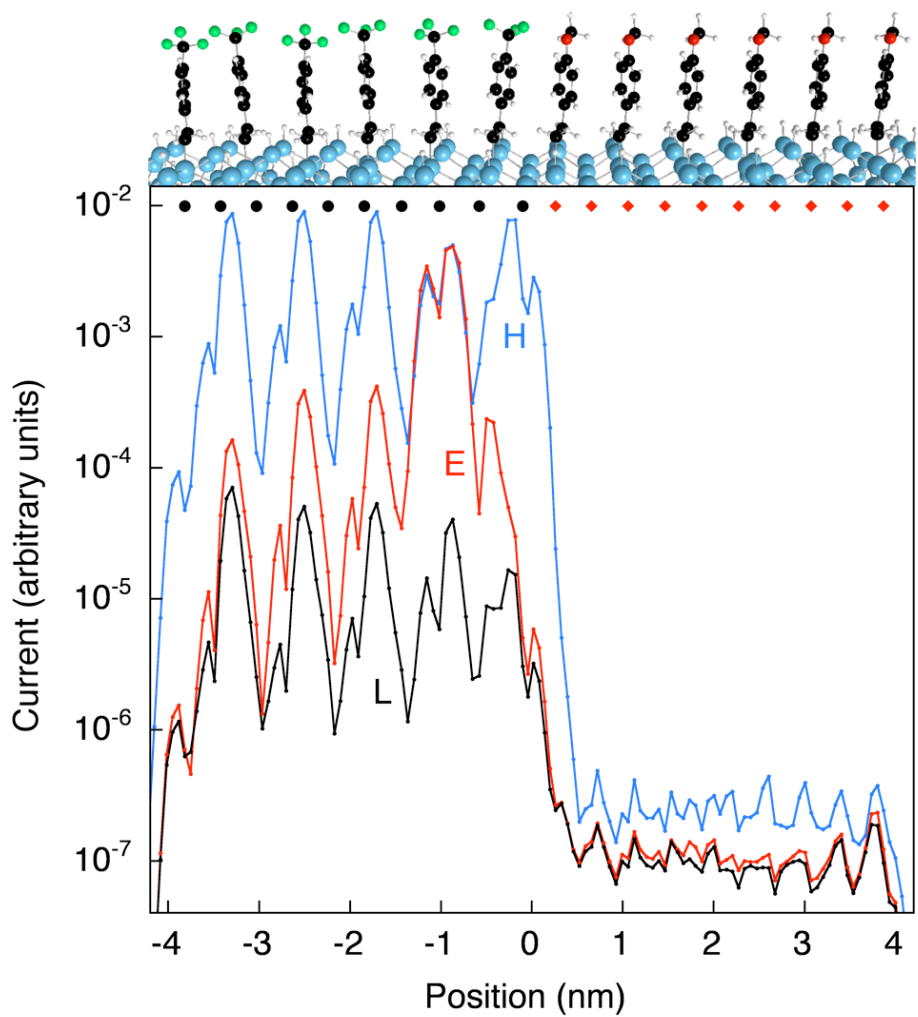}
\caption{\label{Fig_5th}Color online. Calculated current $I$ 
flowing between the tungsten STM tip and a
CF$_3$--styrene/OCH$_3$--styrene molecular chain on silicon at
some positive substrate biases vs. STM tip position
along the chain at constant tip height. 
A side view of a
{\em part} of the molecular chain around the  
CF$_3$--styrene/OCH$_3$--styrene junction is shown at the top of the figure.
Atoms are colored as in Fig.\ref{Fig_4struct}. 
 A top view of the positions of some key atoms is shown 
in Fig. \ref{Fig_8th}(a) together with the trajectory of
the tungsten STM tip for the calculated current profiles. Curve
L (black) is for a very low bias for which the STM 
tip Fermi level is between the bottom
of the silicon conduction band at the Si surface
and the lowest energy state derived from the molecular
LUMOs. Curve E (red) is for a somewhat higher (but still low)
bias for which the STM 
tip Fermi level is just above the lowest energy state 
derived from the molecular LUMOs.
Curve H (blue) is for a still higher bias for which the STM 
tip Fermi level is near the top of
the band of states derived from the molecular LUMO.
Black bullets (red diamonds) at the top
indicate the lateral locations
of the C atoms of the molecular CF$_3$ groups
(O atoms of the OCH$_3$ groups). }
\end{figure*}

\begin{figure*}[b]
\includegraphics[width=1.00\linewidth, clip=true, trim=0.0 200.0 0.0 100.0]{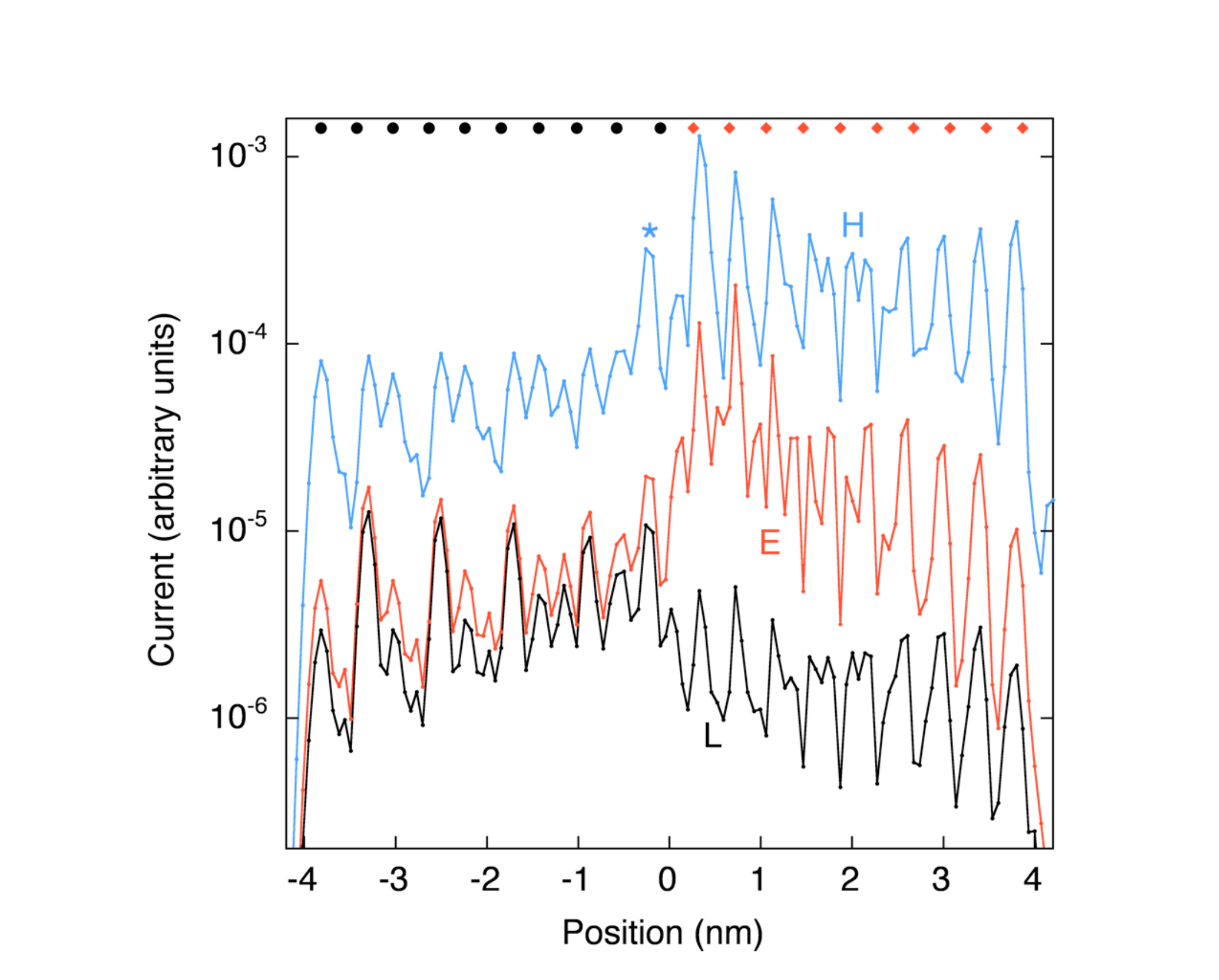}
\caption{\label{Fig_6th}Color online. Calculated current $I$ 
flowing between the tungsten STM tip and a
CF$_3$--styrene/OCH$_3$--styrene molecular chain on silicon at
some negative substrate biases (filled state imaging) vs. STM tip position
along the chain at constant tip height. The geometry
of the chain is that in Fig. \ref {Fig_5th} and Fig. \ref{Fig_8th}(a)
where the trajectory of the STM tip is also shown. Curve
L (black) is for a very low bias for which the STM 
tip Fermi level is between the top
of the silicon valence band at the Si surface
and the highest energy state derived from the molecular
HOMOs. Curve E (red) is for a somewhat higher but still low
bias for which the STM 
tip Fermi level is just below the highest energy state 
derived from the molecular HOMOs.
Curve H (blue) is for a still higher bias for which the STM 
tip Fermi level is just below the bottom of
the band of states derived from the OCH$_3$--styrene 
molecular HOMO.
The black bullets (red diamonds) at the top of the plot
indicate the lateral locations
of the carbon atoms of the molecular CF$_3$ groups
(O atoms of the OCH$_3$ groups).}
\end{figure*}

Fig. \ref{Fig_5th} shows representative results for the
calculated current $I$ 
flowing between the tungsten STM tip and silicon substrate via the 
CF$_3$--styrene/OCH$_3$--styrene chain described in the preceding
paragraph at some positive substrate biases (i.e., for empty molecular state 
imaging) 
vs. STM tip position
along the chain at constant tip height. The corresponding results for 
negative substrate bias (filled molecular state imaging) are shown in Fig. \ref{Fig_6th}
The black bullets (red diamonds) at the top of each plot
indicate the lateral locations
of the carbon atoms of the molecular CF$_3$ groups
(O atoms of the OCH$_3$ groups).

In both figures, curves
L (black) are calculated current profiles
along the molecular chain in the very low bias Regime 1 
introduced in Section \ref{Level Ordering}:
Here the STM tip Fermi level lies between the silicon conduction band edge
and the lowest state of the molecular LUMO band in Fig. \ref{Fig_5th}, and 
between the silicon valence band edge
and the highest state of the molecular HOMO band in Fig. \ref{Fig_6th}. 
The red curves E are in the low bias Regime 2 of Section \ref{Level Ordering}
where the STM tip Fermi level is slightly above (below)
the lowest (highest) molecular LUMO (HOMO) level in Fig. \ref{Fig_5th} (Fig. \ref{Fig_6th}).
The blue curves H are in the higher bias Regime 3 in which the STM tip Fermi level is near the top 
the lowest molecular LUMO band in Fig. \ref{Fig_5th} and just below the bottom of the highest
molecular HOMO band in Fig. \ref{Fig_6th}.

The main features of the current plots  in Figs \ref{Fig_5th} and \ref{Fig_6th} 
can be understood qualitatively by considering the
profiles along the molecular chain of the calculated electrostatic shifts $E_n$ 
of the
atomic orbital energies defined by Eqn.(\ref{electrostics}) that
enter into the present model as discussed in Section \ref{Formalism}. The
relevant information is summarized in Fig. \ref{Fig_7pots} (a).
There the red curve (Ph) shows 
the average of $E_n$ over the six carbon atoms of the benzene ring
of each molecule (the scale is on the left axis) vs. the lateral 
position of the center of the ring along the molecular chain.
Since the HOMO and LUMO of both the CF$_3$--styrene and 
OCH$_3$--styrene are located
principally on the benzene ring, the red curve is an approximate guide as to
how the molecular HOMO and LUMO orbital energies are affected by the
electrostatic potentials due to charge transfer between the atoms of this
system.
The blue dotted vertical line separates the OCH$_3$--styrene part of the
chain on the right from the CF$_3$--styrene part on the left.
The black curve labeled Si (for which the scale is on the right axis) shows 
$E_n$
for the Si atoms to which the molecules bond. 

These electrostatic shifts differ by several
tenths of an eV between the OCH$_3$--styrene and the CF$_3$--styrene parts 
of the molecular chain
due to the different signs and magnitudes of the charge transfer between the 
CF$_3$ and OCH$_3$ substituent groups and the benzene rings of the 
respective molecules.
They result in a further lowering of the HOMO and LUMO of the 
CF$_3$--styrene relative to those of the
OCH$_3$--styrene in addition to that predicted by extended
H{\"u}ckel theory and also a {\em very local} band bending of the silicon 
conduction and valence bands in
the immediate vicinity of the molecular chain; the Si band 
edges are lower in energy
near the 
CF$_3$--styrene chain
than near the  the OCH$_3$--styrene. The horizontal dotted line in Fig. \ref
{Fig_7pots} (a)
indicates $E_n$ for 
surface Si atoms if all molecules are replaced by H atoms bound to the Si; 
the effect of such (passivating) H atoms on the orbital 
energies of the surface Si  atoms is 
intermediate
between those of OCH$_3$--styrene and CF$_3$--styrene molecules
bound to the silicon.

Because of the form of these electrostatic profiles,
when the Si substrate is biased positively relative to the STM
tip, the Fermi level of the metal tip crosses the Si conduction band edge near 
the CF$_3$--styrene
chain  
and the LUMO levels of the CF$_3$--styrene
before doing so at the OCH$_3$--styrene. Thus the
CF$_3$--styrene appears higher than the OCH$_3$--styrene for positive 
substrate bias. This is
seen in the theoretical current plots in Fig. \ref{Fig_5th} and also 
experimentally in Fig. \ref{Fig_2expt}(c)
although, as was discussed in Section \ref{Expt} the contrast between the 
OCH$_3$--styrene and CF$_3$--styrene
depends on the unknown microscopic details of the STM tip. As will be 
discussed in Section \ref{Relationship}
it also depends on the specific conformation of the molecular chain.

The red curve (Ph)  in Fig. \ref{Fig_7pots} (a) has a sharp minimum at the third 
CF$_3$--styrene
molecule from the junction with the OCH$_3$--styrene chain. Thus the 
molecular LUMO level is lowest at
that molecule. Therefore as the (positive) substrate bias increases, the Fermi 
level of the STM tip first crosses a molecular LUMO-derived state in the vicinity 
of that molecule. The prominent peak centered at the third
CF$_3$--styrene molecule in the red current profile E in Fig. 
\ref{Fig_5th} thus
corresponds to the onset of electron injection from the STM tip into the 
molecular LUMO band. There is
no such feature at this molecule in the black curve L in Fig. \ref{Fig_5th} 
(which is at lower bias)
because in that case the STM tip Fermi level is still too far away from the 
molecular LUMO levels for resonant
tunneling through those levels to be important; thus the larger electrostatic 
shift of the third
CF$_3$--styrene molecule's LUMO level does not play a large role here. 
When the bias is large enough for
the STM tip's Fermi level to have risen to the top of the molecular CF$_3$-
styrene LUMO band {\em all} of the molecular LUMO
orbitals throughout the CF$_3$--styrene chain are transmitting resonantly so 
that the difference
between the third CF$_3$--styrene molecule and the other CF$_3$--styrene 
molecules has been largely 
erased in the blue curve H in
Fig.
\ref{Fig_5th}. 

Notice that although the Si band edge is lowered locally around the 
CF$_3$--styrene
molecules (see the black curve in Fig. \ref{Fig_7pots} (a)) so that one may 
expect the Si to appear higher in that
vicinity than elsewhere in empty state STM images, the black curve labelled Si 
in Fig. \ref{Fig_7pots} (a) does
{\em not} show a minimum at the third CF$_3$--styrene molecule from the 
junction, 
confirming that it is the local electrostatic lowering
of the molecular LUMO at that molecule and not the effect of the electric fields 
at the Si that
is responsible for the prominent feature at the third CF$_3$--styrene molecule 
in the red profile E in Fig.
\ref{Fig_5th}. 

At the bias voltages considered in Fig. \ref{Fig_5th} the tip Fermi level is well 
below the
OCH$_3$--styrene LUMO states so that resonant transport via the 
OCH$_3$--styrene LUMO states is not important
and thus there is little contrast between different parts of the OCH$_3$-
styrene chain except possibly at the
highest bias value shown for the OCH$_3$--styrene molecule that is closest to 
the CF$_3$--styrene chain. 

Notice also that in Fig. \ref{Fig_5th} the contrast between the CF$_3$--styrene 
and OCH$_3$--styrene
chains decreases markedly with {\em decreasing} magnitude of the applied bias. 
This is because the 
role of resonant transmission via the molecular levels in the CF$_3$--styrene 
part
of the chain is decreasing rapidly whereas no resonant transmission is 
involved in
conduction via the OCH$_3$--styrene which is therefore less sensitive to the 
magnitude of the applied bias.
This behavior is in
qualitative agreement with the experimental data in Fig. \ref{Fig_2expt}(c) and 
(d).

The qualitative features of the 
theoretical current profiles for negative substrate bias 
in Fig. \ref{Fig_6th} can also be understood by considering the electrostatic
energy profiles in Fig. \ref{Fig_7pots} (a):

The current profile at very low bias
(black curve L, Fig. \ref{Fig_6th}) shows only weak contrast between different 
parts of the
CF$_3$--styrene chain and between different parts of the 
OCH$_3$--styrene chain because
the tip Fermi level is between the silicon valence band edge and the 
molecular
HOMO levels so that resonant tunneling via the molecular HOMO levels (that 
is strongly
modulated by local extrema of the electrostatic potential along
the chain of molecular benzene rings exhibited by the red curve in Fig. \ref
{Fig_7pots} (a)) 
is not occurring.\cite{lobiasexpt}

When the tip Fermi level falls below the highest molecular HOMO-derived 
level (which
is located at and around the second OCH$_3$--styrene molecule from the 
junction where the electronic electrostatic
potential energy [the red curve B in Fig. \ref{Fig_7pots} (a)] has its maximum)
the current develops a strong maximum there because of the onset of 
resonant conduction via the
molecular HOMO in that vicinity. This is why the red curve E of Fig. \ref
{Fig_6th} has a strong maximum
at the second OCH$_3$--styrene molecule from the junction. 

With further 
increase of the bias voltage
the tip Fermi level moves deeper into the molecular HOMO band and 
resonant transport via molecular
states derived from the molecular HOMO becomes possible in other parts of 
the OCH$_3$--styrene chain
as well and thus the peak in the current profile of the OCH$_3$--styrene chain 
becomes less
pronounced, as in the blue curve  H in Fig. \ref{Fig_6th}. Notice also that with 
increasing bias the
boundary between the OCH$_3$--styrene and CF$_3$--styrene chains 
becomes blurred with the higher
current typical of the OCH$_3$--styrene chain extending to the nearest one or 
two CF$_3$--styrene molecules,
as in the feature labelled * in the blue curve H in Fig. \ref{Fig_6th}.

Another feature of Fig. \ref{Fig_6th} is the reversal in the contrast between the
OCH$_3$--styrene and CF$_3$--styrene chains between low and intermediate 
bias: Curve L is higher for the
CF$_3$--styrene while curves E and H are higher for the OCH$_3$--styrene. 
This is possible because the transport 
mechanisms for the OCH$_3$--styrene in the low and
intermediate bias regimes are different: The latter is dominated by resonant 
conduction
via the molecular HOMO levels while the former is not. 
Whether such a contrast reversal with increasing bias 
occurs or not was found in the present 
theoretical study to depend on the
structures of the molecular chains: For the geometry shown at the top of Fig. 
\ref{Fig_5th} (see also Fig. \ref{Fig_4struct}) the OCH$_3$
groups lie especially low relative to the CF$_3$ groups which makes it 
possible for the
CF$_3$--styrene molecules to appear higher than the OCH$_3$--styrene at 
low negative substrate bias. This occurs despite the fact that both the 
molecular HOMO levels and the local Si valence band edge 
are higher in energy on the OCH$_3$--styrene side of the heterowire. For some other
geometries of the OCH$_3$--styrene/CF$_3$--styrene chain to be discussed in 
Section \ref{Relationship} 
the OCH$_3$ groups lie higher
relative to the CF$_3$ groups and no contrast reversal is found theoretically: 
The OCH$_3$--styrene appears
higher than the CF$_3$--styrene even at low negative substrate bias. 
In experiment, contrast reversal between the OCH$_3$--styrene
and CF$_3$--styrene chains with changing bias was observed in 
some runs (as in Fig. \ref{Fig_2expt}) but not in others (as in Fig. \ref{Fig_1expt}), 
most likely due to undetermined differences between the STM tips involved.

A significant difference between the calculated current profiles in 
Figs \ref {Fig_5th} and  \ref {Fig_6th}
and our experimental STM height profiles is that the latter
do not show the quasiperiodic modulation with a period of roughly
two molecular spacings seen in the calculated current profiles for the
CF$_3$--styrene part of the chain.
This difference can be understood as follows:
This quasi-periodic modulation of the calculated CF$_3$--styrene chain 
current
profile is a geometrical effect due to the differing orientations
of the CF$_3$ groups and also the associated differing tilts of the molecules
along the CF$_3$--styrene chain described near the end  
of Section
\ref{Structure}. Therefore it is
reasonable to expect this modulation not to be visible in STM
images taken at room temperature where the CF$_3$ groups are
rotating rapidly about their axes and the local phase of the spatial current 
modulation is
therefore fluctuating rapidly so that only an average over many
structures with different phases is observed on the time scale
on which the STM scans are recorded, and thus a period equal to the
molecular spacing is observed.

\subsection{The current peaks near the 
CF$_3$--styrene/OCH$_3$--styrene interface }
\label{Interface}

The prominent peaks in the red current profiles E in 
Figs \ref {Fig_5th} and  \ref {Fig_6th} near the interface between the 
CF$_3$--styrene and OCH$_3$--styrene molecular
chains have the following important characteristics: 
The peak for positive substrate bias in Fig. \ref {Fig_5th} is narrow, just one
molecule wide and occurs on the CF$_3$--styrene side of the junction. 
The peak at negative substrate bias in Fig. \ref {Fig_6th} is broader, extending
over a few molecules and occurs on the OCH$_3$--styrene side of the junction.
These properties of the interfacial peak at negative substrate bias are consistent
with the experimental data presented in Section \ref{Expt}. Furthermore the interfacial 
peak at negative substrate bias was seen experimentally at low bias in all of the
many CF$_3$--styrene/OCH$_3$--styrene heterojunctions 
that we imaged and persisted when
the STM tip changed spontaneously in the course of the imaging.
By contrast, in empty state imaging (i.e., for positive substrate bias), 
no discernible height enhancement at the 
interface was usually observed. 

The above discussion in Section \ref{Positive} identified the interfacial 
peaks in the calculated current profiles
as being due to a maximum or minimum of the electrostatic 
contribution to the molecular HOMO or LUMO energy along
the molecular chain occurring at the location of the peak. 
However, in order to
understand the physics underlying the differences between 
the predicted properties of the peaks at positive and negative 
substrate bias and the similarities and differences between  
the predictions and experiment 
it is necessary to examine how these features of the current 
profiles depend on the structure of the molecular chain. 
This will be explored in the next section.

\section{Relationship between structure, electrostatics and current profiles}
\label{Relationship}

Several significant aspects of the structures of the experimentally realized 
molecular chains such as whether the
molecules are tethered to the Si in the T or D orientation (as defined in 
Section
\ref{Geom}) and whether the OCH$_3$ groups are T or D oriented, are not 
obvious from a direct inspection of
the experimental STM images. It is also not clear theoretically which types of 
these structures should be
favoured by the kinetics of the growth process or by energetics at room 
temperature. Thus it is desirable to study
a variety of plausible relaxed molecular chain geometries theoretically and to 
clarify how their structures
relate to their electrostatic and current profiles. This is done in the present 
Section and the results
presented here also elucidate the mechanisms underlying the phenomena 
described in Sections \ref{Expt} and \ref{Case}.
Potential energy profiles were calculated for many relaxed chain geometries 
and a few representative examples will be discussed below,
along with possible experimental implications. 

\subsection{Notation}
\label{Notation}

The important features of these molecular chain geometries are 
summarized in Fig. \ref{Fig_8th} where projections of the positions of some key atoms 
of the molecular chains onto the Si (001) plane are shown: Each panel in 
Fig. \ref{Fig_8th} shows two rows of Si surface atoms 
belonging to {\em different} Si dimer rows. In each case
the molecules are tethered to
the lower row of Si atoms. The C atoms that bond to
the Si atoms are orange. The C atoms that bond to the C atoms that bond to
the Si atoms are violet. The C atoms of the CF$_3$ and OCH$_3$ groups are 
black.
The O atoms are red and the F atoms are green. The C atoms belonging to 
the benzene 
rings that bond to O atoms or to C atoms not belonging to the benzene rings 
are 
white. The other other C atoms of the benzene rings, the H atoms and the 
other Si
atoms are not shown for clarity. The dashed line shows the path of the tip 
atom of the STM for the theoretical STM current profiles for each structure
that are presented in this paper.

The structure considered in Section \ref{Case} will henceforth
be referred to as ``Structure (a)" as per its label in Fig.
\ref{Fig_8th}. The other structures to be discussed here will be referred to 
similarly, according to their
labelling in Fig.
\ref{Fig_8th}. Each of these structures consists of 20 molecules taken from the 
central region of a relaxed 40
molecule chain without further relaxation, as described at the end of Section 
\ref{Formalism}.

\subsection{Structure (b)}
\label{Structure (b)}

In Structure (b), as in Structure (a), the OCH$_3$--styrene molecules are 
tethered to the Si in the T
orientation. But the OCH$_3$ groups are approximately reversed relative to 
Structure (a) and are now in
the D orientation (defined in Section \ref{Geom}). 
In Structure (b) the tethering of the CF$_3$--styrene 
molecules to the Si alternates between
T and D and the orientation of the F atoms in the CF$_3$ groups also 
alternates from molecule to molecule.
However since this is a relaxed structure neither the OCH$_3$--styrene chain 
nor the CF$_3$--styrene chain has a
truly periodic geometry. 

\subsubsection{Electrostatic and current profiles}
\label{ecprofiles}

The electrostatic potential energy profiles along the 
chain of benzene rings and
along the row of Si atoms to which the molecules of Structure (b) bond are the 
red and black curves in Fig.
\ref{Fig_7pots}(b), resp. 
Some representative current profiles for
Structure (b) calculated at constant tip height along the dashed line in Fig. \ref
{Fig_8th}(b) are shown in
Fig. \ref{Fig_9th} together with a side view of a part of this molecular chain. 
The qualitative behavior of the calculated current profiles in 
Fig. \ref{Fig_9th} can be understood by considering 
the features of the potential energy profile along the chain of benzene 
rings in Fig. \ref{Fig_7pots}(b) and applying reasoning analogous
to that in Section \ref{Positive}: The current profile under a moderately low positive 
substrate bias for which the STM tip
Fermi level is just above the lowest level of the molecular LUMO band (the 
violet curve E$'$) has a strong
maximum one molecule wide at the CF$_3$--styrene molecule next to the 
OCH$_3$--styrene chain where the
benzene ring potential energy profile has its sharp minimum. The current 
profile under a moderately low 
negative substrate bias for which the STM tip
Fermi level is just below the highest level of the molecular HOMO band (the 
red curve E) rises
gradually to a broad maximum at the end of the OCH$_3$--styrene chain 
{\em remote from} the CF$_3$--styrene
where the benzene ring potential energy profile has its broad maximum. 
Unlike in
Fig.\ref{Fig_6th}, in this case there is no contrast
reversal between the OCH$_3$--styrene and CF$_3$--styrene chains with 
increasing negative substrate bias; the
OCH$_3$--styrene is higher in both the very low bias profile L and the 
moderately low bias profile E. This is
because the OCH$_3$ group is located higher relative to the CF$_3$ group 
in Structure (b) than in Structure (a). 

\subsubsection{Significance of the results}
\label{Significance}

\begin{figure*}[b]
\includegraphics[width=1.00\linewidth, clip=true, trim=0.0 170.0 0.0 0.0]{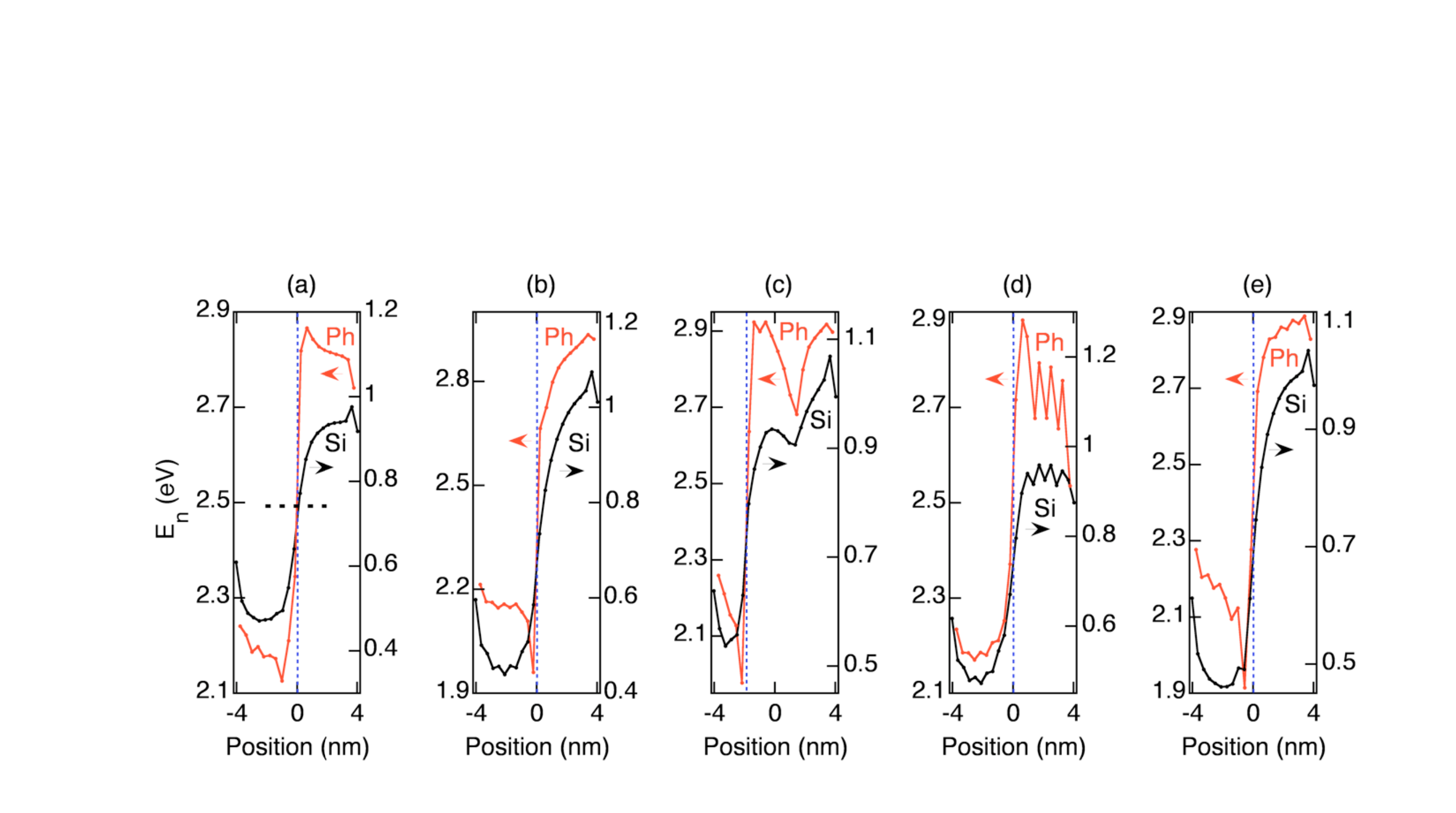}
\caption{\label{Fig_7pots}
Calculated electrostatic shifts of atomic orbital energies
$E_n$ defined by Eqn. (\ref{electrostics}) vs. lateral position
of the orbital along the molecular chain for various molecular
chain geometries. Red curves (Ph) show 
the average of $E_n$ over the six carbon atoms of the benzene ring
of each molecule (scale on the left axis), black curves labelled Si (right axis) 
show $E_n$ for the Si atoms to which the molecules
bond. Blue vertical dashed lines
separate the CF$_3$--styrene part of the chain on the left from the OCH$_3$-
styrene on the
right. Case (a) is for the molecular chain geometry discussed
in Section \ref{Case}; the calculated current
profiles for this geometry are shown in Figs \ref{Fig_5th} and \ref{Fig_6th}.
The horizontal black dotted line in (a) indicates $E_n$ for 
surface Si atoms if the molecules are replaced by H atoms. 
The distinguishing structural features of the molecular
chains with the electrostatic profiles shown in panels (a)-(e) are presented
in Fig. \ref{Fig_8th}(a)-(e), respectively.}
\end{figure*}

\begin{figure*}
\includegraphics[width=1.00\linewidth, clip=true, trim=0.0 150.0 0.0 0.0]{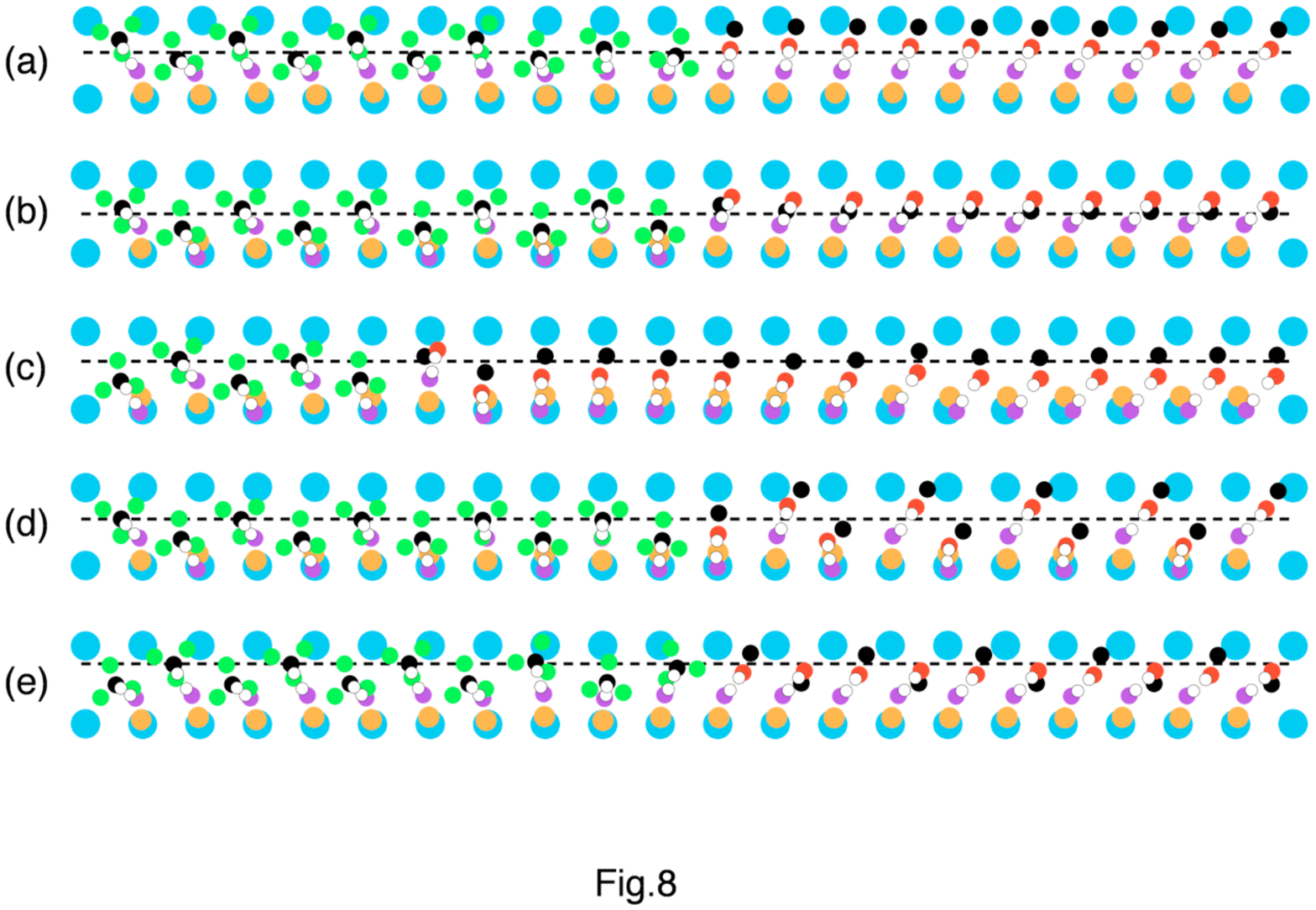}
\caption{\label{Fig_8th}Atomic positions projected onto the Si (001) 
surface for some CF$_3$--styrene/OCH$_3$--styrene
molecular chains on silicon studied theoretically in this work. Each panel 
shows two rows of Si atoms 
belonging to {\em different} Si dimer rows. The molecules are tethered to
the lower row of Si atoms in each panel. The C atoms that bond to
the Si atoms are orange. The C atoms that bond to the C atoms that bond to
the Si atoms are violet. The C atoms of the CF$_3$ and OCH$_3$ groups are 
black.
The O atoms are red and the F atoms are green. The C atoms belonging to 
the benzene 
rings that bond to O atoms or to C atoms not belonging to the benzene rings 
are 
white. The other other C atoms of the benzene rings, the H atoms and the 
other Si
atoms are not shown for clarity. The dashed lines show the paths of 
the apex 
atom
of the STM tip for the theoretical STM current profiles presented in this paper.
(a) is the structure for Fig. \ref{Fig_5th} and \ref{Fig_6th}. (b),(c),(d),(e) 
correspond 
to Figs \ref{Fig_9th},\ref{Fig_10th},\ref{Fig_11th},\ref{Fig_12th} respectively. }
\end{figure*}

\begin{figure*}
\includegraphics[width=0.90\linewidth, clip=true, trim=0.0 100.0 0.0 50.0]{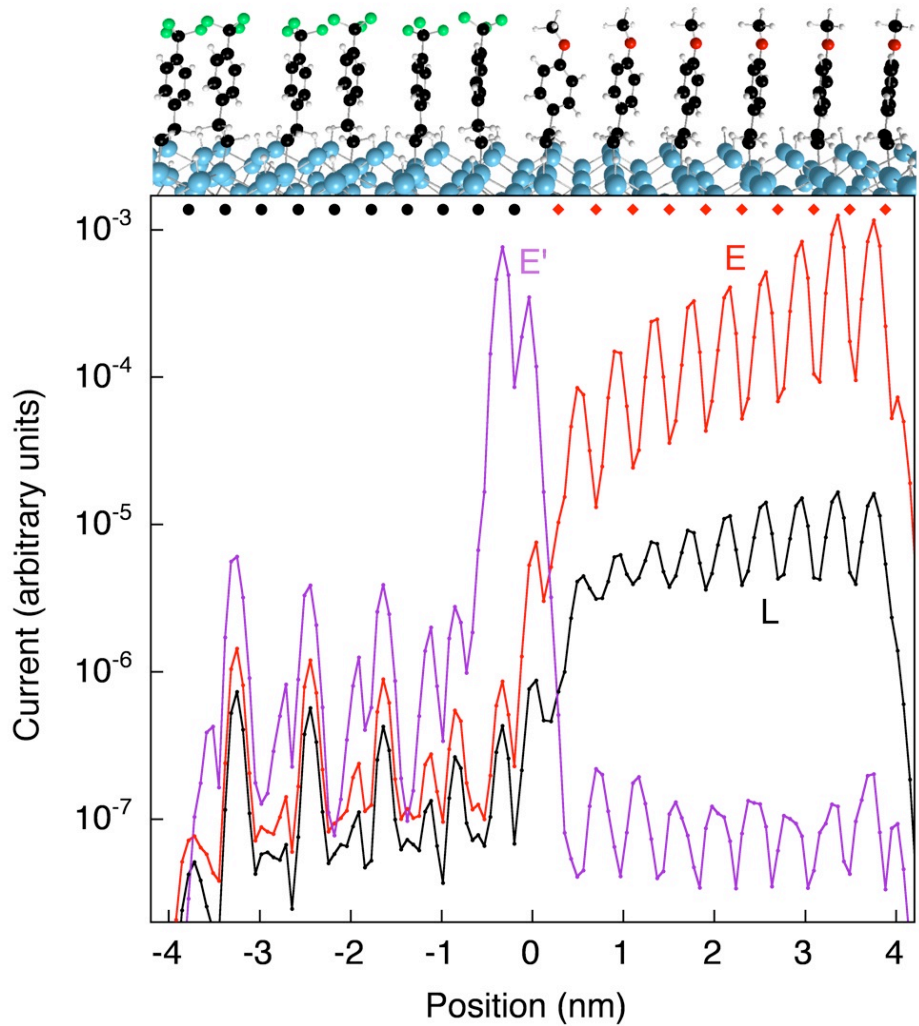}
\caption{\label{Fig_9th}Calculated current $I$ 
flowing between the tungsten STM tip and a
CF$_3$--styrene/OCH$_3$--styrene molecular chain on silicon vs. STM tip 
position
along the chain at constant tip height for the molecular geometry shown at the
top of the figure and in Fig. \ref{Fig_8th}(b)
where the trajectory of the STM tip is also shown.
Notation as in Fig.\ref{Fig_5th}. Curve L (black) is for a very low negative 
substrate bias for which the STM 
tip Fermi level is between the top
of the silicon valence band at the Si surface
and the highest energy state derived from the molecular
HOMOs. Curve E (red) is for a somewhat higher but still low negative 
substrate bias for which the STM 
tip Fermi level is just below the highest energy state 
derived from the molecular HOMOs.
Curve E$'$ (violet) is for a low positive substrate bias for which the STM 
tip Fermi level is just above the lowest state derived from the molecular 
LUMO.
}
\end{figure*}

The two main differences between the electrostatic profiles for  Structure
(b) and those for Structure
(a) (shown in Fig. \ref{Fig_7pots}(b) and Fig. \ref{Fig_7pots}(a)) are: 

1. The minimum of the benzene ring electronic 
potential energy profile (while remaining 
very sharp) has shifted from the third CF$_3$--styrene molecule from the 
heterojunction in Fig.
\ref{Fig_7pots}(a)  to the CF$_3$--styrene molecule next to the heterojunction
in Fig. \ref{Fig_7pots}(b).

2. The maximum of the benzene ring potential energy profile while remaining 
broad has shifted from the vicinity of the heterojunction (on the 
OCH$_3$--styrene side) in Fig. \ref{Fig_7pots}(a) to the vicinity of the end
of the OCH$_3$--styrene chain that is remote from the heterojunction
in in Fig. \ref{Fig_7pots}(b).

Systematic studies of various structures and their potential energy profiles 
showed the reversal of the slope
of the potential energy profile along most the chain of benzene rings of the 
OCH$_3$--styrene chain from
Structure (a) to Structure (b)  in Fig. \ref{Fig_7pots} should be
attributed {\em not} to the OCH$_3$ groups being switched from the T to the D 
orientation {\em per se}, but 
primarily to the fact that the {\em projection onto the axis of the molecular 
chain} of the dipole moment associated 
{\em internal} charge structures of the OCH$_3$ groups (negative O and 
positive CH$_3$) has reversed direction from Structure (a) to Structure
(b) as is evident from the geometries of the two structures in Fig. \ref{Fig_8th}. 
This reversal of the
charge structure along the molecular axis should lead to a reversal 
of the slope of the potential
energy profile and this is what is in fact seen in Fig. \ref{Fig_7pots}. 

In this way the change
in the geometrical structure of the OCH$_3$--styrene chain has
resulted in the shift of the potential energy maximum (and STM tip current 
maximum at moderately low negative substrate bias) 
from the vicinity of the heterojunction to the vicinity of the far end  
of the OCH$_3$--styrene chain.

An important point to note is that because of the geometrical structure
of the OCH$_3$--styrene molecules and the way in which they bond to
the silicon  the  internal
dipole moment of the OCH$_3$ group has a strong 
component {\em parallel} to the silicon 
surface and the OCH$_3$
groups of all of the OCH$_3$--styrene molecules can be aligned in such a way 
that the
projections of the OCH$_3$ groups of {\em all} of these molecules onto the 
axis of the molecular
chain have the same sign.
This results in a gradual buildup of the electrostatic potential
along the OCH$_3$--styrene chain and a {\em broad} electrostatic 
potential energy peak near one of its ends as in Fig. \ref{Fig_7pots}(a) or (b),
and consequently a broad current peak under moderately low negative
substrate bias, consistent with the experimental results
presented in Section \ref{Expt}.

By contrast, because the geometrical orientation of the
CF$_3$ group (see Fig. \ref{Fig_4struct}) 
 and the arrangement of charges within it are different
than for the OCH$_3$ group, 
and because of the repulsion between the F atoms on 
different molecules  (see the 
end of Section \ref{Structure}),
such a well ordered arrangement of
dipole projections onto the axis of the molecular chain does not occur
for the CF$_3$--styrene chain. Thus the electrostatic potential energy
minimum on the CF$_3$--styrene side of the heterojunction is a much
more local phenomenon. Its location is determined
by the orientations of a few molecules surrounding it and by the
fringing electrostatic field of the OCH$_3$--styrene chain that extends
across the heterojunction into the  CF$_3$--styrene chain. 
Thus the electrostatic potential energy
minimum on the CF$_3$--styrene side of the heterojunction is very
narrow and its location is sensitive to the details of 
the local molecular geometry.
It is therefore reasonable to expect any profile peak associated with
it in STM imaging at room temperature (at positive substrate bias), if
present at all,  to be more strongly impacted by thermal
fluctuations in the molecular geometry and therefore
less readily observed than the corresponding peak
at negative substrate bias on the OCH$_3$--styrene side of 
the molecular chain. This is consistent with the differences
between the experimental results obtained under positive
and negative substrate bias that are
presented in Section \ref{Expt}.

Experimentally, height enhancement was 
observed in the STM images towards the end of the
OCH$_3$--styrene chain that is far from the junction with
the CF$_3$--styrene at low negative substrate bias as can be
seen, for example, in Fig. \ref{Fig_1expt}(d). Because
in this work the growth of the OCH$_3$--styrene chain
followed the growth of the CF$_3$--styrene,  
a charged dangling
bond is normally expected to be present at the end of the
OCH$_3$--styrene chain (remote from the 
OCH$_3$--styrene/CF$_3$--styrene junction) where the
growth of the molecular chain terminated, as in
previous experimental work on styrene chains
on silicon.\cite{Lopinski,Nature} The electrostatic
potential due to this charged dangling bond 
gives rise to enhancement of the height  profile 
of the molecular chain in its vicinity in STM experiments
as is discussed in Ref.\onlinecite{Nature} If 
alignment of OCH$_3$--styrene molecular dipoles as in
Structure (b) was occurring in the present experiments,
the associated electrostatic profiles such
as in Fig. \ref{Fig_7pots}(b) would {\em also} 
contribute to enhanced current
near the end of the molecular chain as in curve E in Fig.
\ref{Fig_9th}, and hence to the enhanced height
profile there in constant current STM experiments
as in Fig. \ref{Fig_1expt}(d).
Determining experimentally whether such a molecular 
dipole contribution to the height profile in vicinity of
the dangling bond is present or not is in principle
possible by passivating the dangling bond through 
the addition of a H atom and observing whether any
height enhancement remains afterwards near the 
end of OCH$_3$--styrene chain. Carrying out such a test
was beyond the scope of the present experimental work. 
However, the fact that height enhancement was {\em
consistently} observed experimentally near the
OCH$_3$--styrene/CF$_3$--styrene heterojunction at low
negative substrate bias suggests that OCH$_3$--styrene
chains with dipole fields similar to that arising from
Structure (a) rather than Structure (b) in Fig. 8 
played the dominant role in these experiments.  

\subsection{Structure (c)}
\label{Structure (c)}

Structure (c) consists of 5 CF$_3$--styrene and 15 OCH$_3$--styrene 
molecules. The geometry of the
CF$_3$--styrene molecules and of the OCH$_3$--styrene molecule that is next 
to the CF$_3$--styrene is similar to
that of the corresponding molecules in Structure (b). However,
as can be seen in Fig. \ref{Fig_8th}(c), the other OCH$_3$--styrene molecules 
in Structure (c) have a
different geometry: they are tethered to the Si in the D orientation and their 
OCH$_3$
groups are in the T orientation. Between the 8th and 9th OCH$_3$--styrene 
the molecules from the junction
with the CF$_3$--styrene there is another abrupt change in the molecular 
geometry: Although the molecules to the
right and left of this dislocation are both D-tethered and the OCH$_3$'s are T 
oriented, the molecules
to the right of the fault tilt more to the right and their OCH$_3$ groups are 
oriented nearly 
perpendicularly to the axis of the molecular chain.

This is reflected in the electrostatic potential energy profile along the chain 
that is
plotted in Fig. \ref{Fig_7pots}(c): Since the molecular structure at the junction 
is similar to that
for Structure (b) the potential profile on the benzene rings is also similar 
there, i.e., there is a sharp 
potential energy minimum at the CF$_3$--styrene molecule at the junction. 
Since for OCH$_3$--styrene molecules 
2-8 from the junction the orientation of the O and two C atoms bonded to it is 
similar to that for the 
OCH$_3$--styrene molecules in Structure (a) the orientation of the relevant 
dipoles on that part of the chain
is also similar and thus the potential energy profile is also qualitatively similar: 
There is a broad
potential energy maximum on the OCH$_3$--styrene benzene rings peaked 
near the junction with the
CF$_3$--styrene. However, the different orientation of the charged groups of 
the of the 9th-15th OCH$_3$--styrene molecules from the junction results in a 
reversal of the slope of the potential profile and the electronic potential energy  
on the OCH$_3$--styrene benzene rings rises to another maximum near the 
right hand
end of the molecular chain. 

\begin{figure*}
\includegraphics[width=0.90\linewidth, clip=true, trim=0.0 150.0 0.0 50.0]{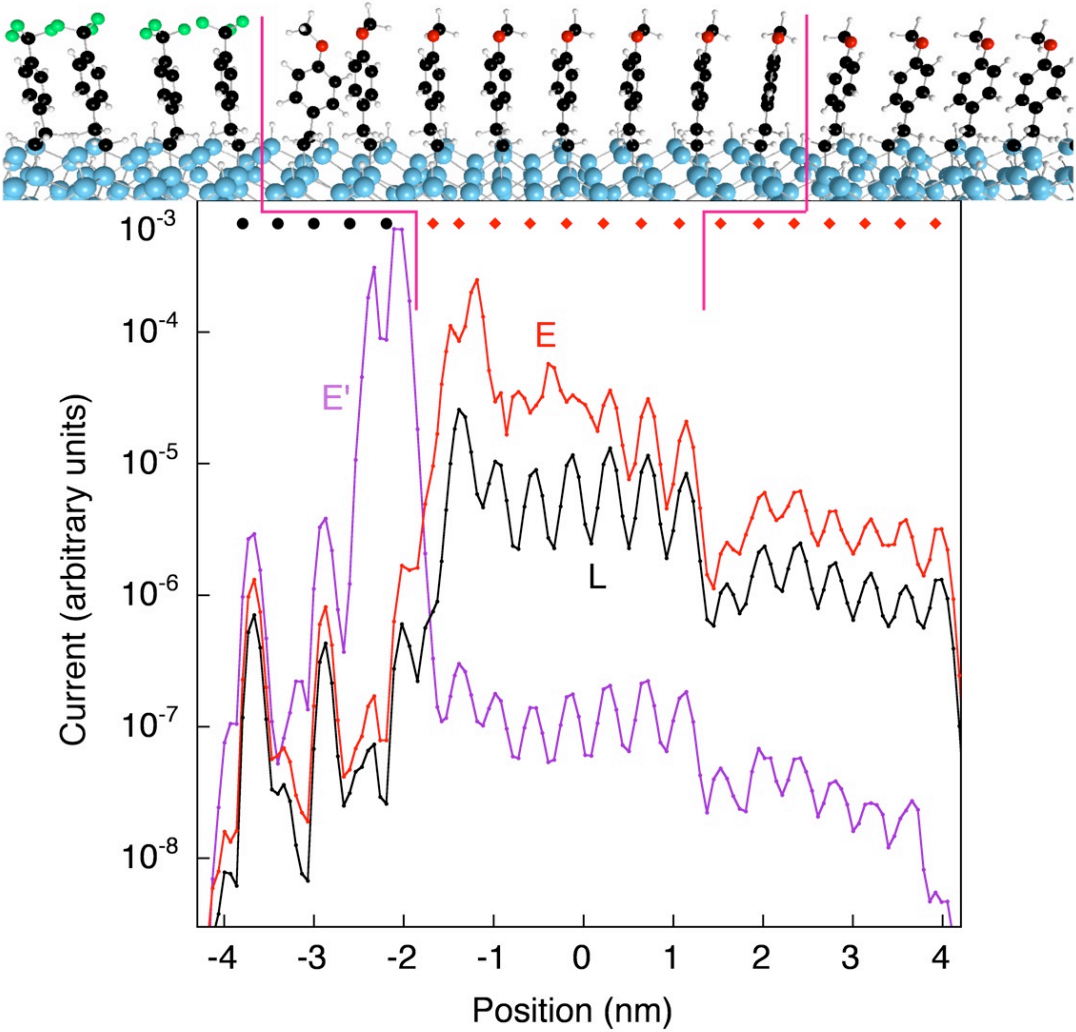}
\caption{\label{Fig_10th}Calculated current $I$ 
flowing between the tungsten STM tip and a
CF$_3$--styrene/OCH$_3$--styrene molecular chain on silicon vs. STM tip 
position
along the chain at constant tip height for the molecular geometry shown at the
top of the figure 
and in Fig. \ref{Fig_8th}(c)
where the trajectory of the STM tip is also shown.
The magenta lines mark the locations of the
CF$_3$--styrene/OCH$_3$--styrene
interface (left) and of the dislocation in the OCH$_3$--styrene chain
(right) in the image at the top of the figure and in the current plots below.
Notation as in Fig.5: Curve L (black) is for a very low negative substrate bias 
for which the STM 
tip Fermi level is between the top
of the silicon valence band at the Si surface
and the highest energy state derived from the molecular
HOMOs. Curve E (red) is for a higher but still low 
negative substrate bias bias for which the STM 
tip Fermi level is just below the highest energy state 
derived from the molecular HOMOs.
Curve E$'$ (violet) is for a positive substrate bias for which the STM 
tip Fermi level is just above the lowest state derived from the molecular 
LUMOs.
}
\end{figure*}

The calculated STM current profiles for Structure (c) 
shown in Fig. \ref{Fig_10th}
are again consistent with the potential energy profile along the chain of 
benzene rings: At moderately low
positive substrate bias (violet curve E$'$) the current profile has a sharp 
maximum one molecule wide at the
CF$_3$--styrene molecule where the potential energy minimum along the 
molecular chain is located. At moderately low
negative substrate bias (red curve E) there is a broad current maximum 
peaked on the OCH$_3$--styrene side of
the junction where the potential energy has one of its maxima. From there the 
current falls off to a minimum
near the middle of the OCH$_3$--styrene chain where the potential energy 
curve also has its minimum. To the
right of this minimum both the potential energy and current rise again but the 
current does not rise as high
as near the junction and then begins to fall off again. This difference is due to 
the fact that for the red
curve E the electron transmission through the molecules has not quite 
reached resonance for the
OCH$_3$--styrene HOMOs at the right end of the OCH$_3$--styrene chain 
(although it has done this for 
the OCH$_3$--styrene near the junction) and also because the 
OCH$_3$--styrene molecules are tilted more
at the right end of the chain than at the left. This geometrical difference also 
shows up as
a step in the other current profiles in Fig. \ref{Fig_10th} at the dislocation in the 
OCH$_3$--styrene chain. 

For Structure (c) there is again no reversal of the contrast between 
OCH$_3$--styrene and CF$_3$--styrene
chains with increasing bias.

\subsection{Structure (d)}
\label{Structure (d)}

In Structure (d) the geometry of the CF$_3$--styrene chain is broadly similar to 
those of
Structures (b) and (c). However, in the OCH$_3$--styrene chain the tethering 
to the Si  
alternates between D and T. The orientation of the OCH$_3$ groups is all T 
and most of the
relevant OC dipoles are oriented roughly as for Structure (a). Thus one should 
expect
a potential energy profile along the chain of OCH$_3$--styrene benzene rings 
somewhat
similar to that for Structure (a). Indeed the red curve in Fig. \ref{Fig_7pots}(d) 
does 
have its maximum in the OCH$_3$--styrene chain near the junction with the 
CF$_3$--styrene
and it declines overall to the right, but the decline is modulated by strong 
oscillations
due to the different tethering of alternate OCH$_3$--styrene molecules to the 
Si.
The potential energy profile along the chain of benzene rings on the 
CF$_3$--styrene chain
of Structure (d) is quite different than that for Structures (b) and (c): There is
no potential energy minimum on the CF$_3$--styrene next to the junction but 
instead
a shallow minimum near the middle of the CF$_3$--styrene chain. 

\begin{figure*}
\includegraphics[width=0.90\linewidth, clip=true, trim=0.0 150.0 0.0 100.0]{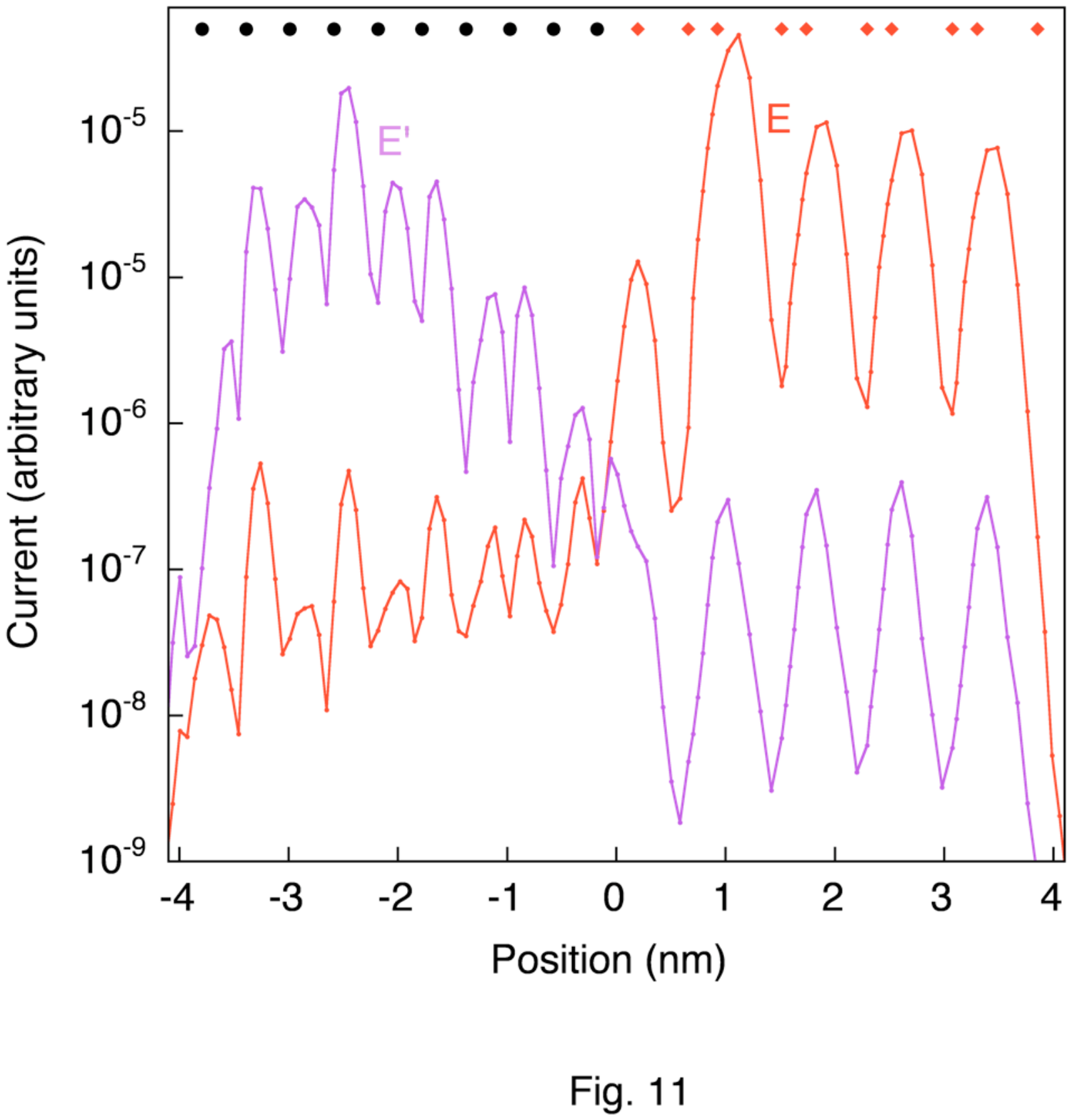}
\caption{\label{Fig_11th}Calculated current $I$ 
flowing between the tungsten STM tip and a
CF$_3$--styrene/OCH$_3$--styrene molecular chain on silicon vs. STM tip 
position
along the chain at constant tip height for the molecular geometry shown in Fig. 
\ref{Fig_8th}(d)
where the trajectory of the STM tip is also shown.
Curve E (red) is for a low negative substrate bias bias for which the STM 
tip Fermi level is just below the highest energy state 
derived from the molecular HOMOs.
Curve E$'$ (violet) is for a positive substrate bias for which the STM 
tip Fermi level is just above the lowest state derived from the molecular 
LUMOs.
}
\end{figure*}

The current profiles for Structure (d) are shown in Fig. \ref{Fig_11th} for 
moderately low positive substrate bias
(violet curve E$'$) and moderately low negative substrate bias
(red curve E). Again the maxima of these profiles are near the lowest and 
highest points of the   
potential energy profile along the chain of benzene rings, respectively. But in 
this case the
potential energy peak in the OCH$_3$--styrene chain is less prominent and 
narrower than in the the
previously discussed cases due to the less uniform structure of the array of 
dipoles. The broad
potential energy minimum in the CF$_3$--styrene results in a broader current 
peak at positive substrate bias
than in the cases discussed earlier. 

This 
illustrates further
the strong sensitivity of the CF$_3$--styrene chain's potential 
minima
to the precise details of the conformation of a single molecule or a few 
molecules: In Structure (d) the
OCH$_3$--styrene molecule at the junction of the chains is D-tethered and the 
OCH$_3$ group is T oriented while
the reverse is true for Structures (b) and (c) while the CF$_3$ groups are 
rotated slightly differently and
CF$_3$--styrene at the junction tilts somewhat differently.

\subsection{Structure (e)}
\label{Structure (e)}

Structure (e) is similar to Structure (a) except that now the orientations of the 
OCH$_3$ groups alternate
between T and D and the CF$_3$ groups have rotated about their axes 
through substantial angles. Because
of the alternating orientations of the OCH$_3$ groups the electric fields due to 
the different OCH$_3$--styrene
molecules are not aligned and produce a potential energy profile in Fig. \ref
{Fig_7pots}(e) with a very broad
maximum near the end of the chain remote from the junction. The different 
orientations of the 
CF$_3$ groups than in Structure (a) result in a sharp potential energy 
minimum at the second CF$_3$--styrene
molecule from the junction.

\begin{figure*}
\includegraphics[width=0.90\linewidth, clip=true, trim=0.0 150.0 0.0 100.0]{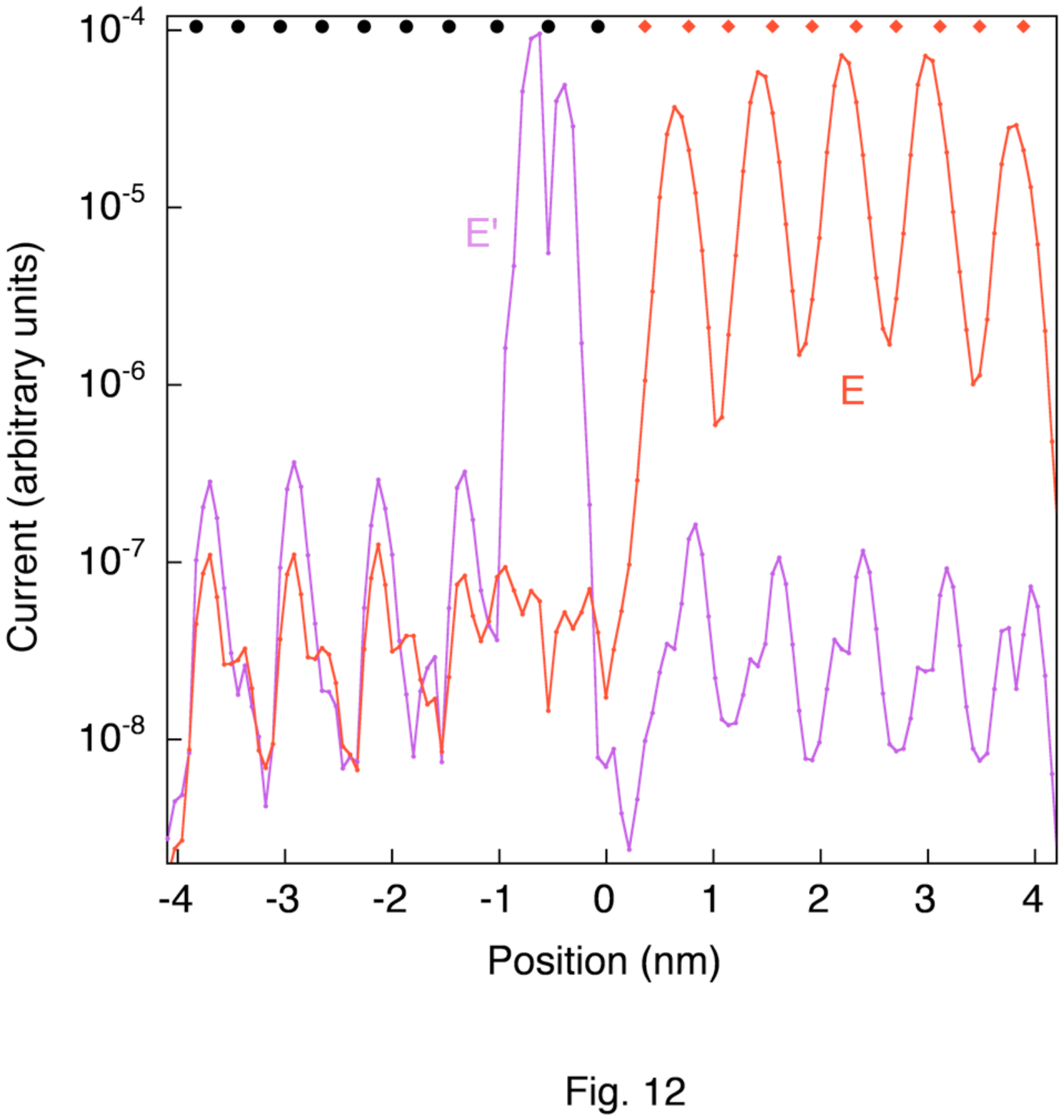}
\caption{\label{Fig_12th}Calculated current $I$ 
flowing between the tungsten STM tip and a
CF$_3$--styrene/OCH$_3$--styrene molecular chain on silicon vs. STM tip 
position
along the chain at constant tip height for the molecular geometry shown in Fig. 
\ref{Fig_8th}(e)
where the trajectory of the STM tip is also shown.
Curve E (red) is for a low negative substrate bias for which the STM 
tip Fermi level is just below the highest energy state 
derived from the molecular HOMOs.
Curve E$'$ (violet) is for a positive substrate bias for which the STM 
tip Fermi level is just above the lowest state derived from the molecular 
LUMOs.
}
\end{figure*}

The current profiles for Structure (e) are shown in Fig. \ref{Fig_12th} for 
moderately low positive substrate bias
(violet curve E$'$) and moderately low negative substrate bias (red curve E). 
As expected for such a potential
energy profile we see a strong narrow current peak on the second
CF$_3$--styrene
molecule from the junction. The broad potential energy maximum on the  
OCH$_3$--styrene chain produces a
broad current maximum peaked towards the end of the OCH$_3$--styrene 
chain that is remote from the junction at 
intermediate negative substrate bias.

For both Structure (e) and Structure (d) the alternating geometries along the 
OCH$_3$--styrene chains
result in a strong modulation of the current with a period of 2 molecular 
spacings. Since this
modulation is not seen experimentally, either these structures are not realized 
in the experiments
or the experiments are averaging over geometries for which the modulation 
occurs with different phases
due to thermal fluctuations of the molecular geometries at room temperature.  
    
\section{Electrostatic vs. Electronic Molecule-Molecule Coupling}
\label{EvE}

For comparison, similar calculations to those described above were also 
carried out for Structure (a)

(i) {\em omitting} the environmental  
electrostatic shifts of atomic 
orbital energies
$E_n$ defined by Eqn. \ref{electrostics}. No enhancement of the current near 
the heterojunction
(such as in curves E in Fig. \ref{Fig_5th} and \ref{Fig_6th}) was found in those 
calculations. 

(ii)Transport calculations were also carried out including
the electrostatic shifts $E_n$ but with all Hamiltonian matrix elements and 
basis function overlaps
responsible for direct electronic hopping from molecule to molecule switched 
off. These calculations produced
very similar current profiles along the molecular chain to those in Fig. \ref
{Fig_5th} and \ref{Fig_6th}, including the interface current peaks
and even the feature labelled $\star$  in the blue curve H in Fig. \ref{Fig_6th}. 
However these features appeared at somewhat different values of the bias 
voltages because the
molecular HOMO and LUMO bands are narrower (as expected) when the 
inter-molecule Hamiltonian matrix elements and basis state overlaps are 
turned off.
 
These results confirm that the enhanced current features near the interface 
in Fig. \ref{Fig_5th} and \ref{Fig_6th} result from electrostatic fields established 
by the polar molecules.  These results therefore suggest an electrostatic origin 
for the filled-state interfacial enhancement resolved experimentally in 
Fig. \ref{Fig_1expt} and \ref{Fig_2expt} at low bias.  (Significantly, no such 
interfacial current enhancement was observed experimentally\cite
{stymeth} in 
styrene/methylstyrene heterostructures on Si where molecular electric 
fields are expected to be much weaker because of the weaker charge 
transfer within those molecules.) 

Furthermore,
although electronic hopping from molecule to molecule may be occurring 
quite efficiently in substituted styrene chains on Si(100) as is discussed in Ref. 
\onlinecite{stymeth}, these calculations demonstrate the hopping not to be 
necessary for the occurrence of the interfacial features reported in the
present work. 

\section{Modeling Single-Triple row CF$_3$--styrene Heterostructures
and their influence on conduction through the underlying Silicon}
\label{Single-Triple}

The main features of the calculated current 
profiles of the molecular heterostructures
modeled theoretically in the preceding sections derive directly from
the electronic structures of the molecules themselves modulated
by the electrostatic fields due to other nearby molecules.
This however is not the case for the most striking feature of the experimental
data shown Fig. \ref{Fig_3expt}(d) and (e) where at low negative substrate
bias the electric current passing through a 
triple row of CF$_3$--styrene molecules is strongly depressed relative to
that passing through the nearby single row of CF$_3$--styrene molecules.
Our theoretical findings presented below suggest that the 
influence of the adsorbed CF$_3$--styrene
molecules on the
electrostatic potential {\em in the underlying silicon} may be 
responsible 
for this phenomenon.

\begin{figure*}
\includegraphics[width=1.00\linewidth, clip=true, trim=0.0 120.0 0.0 0.0]{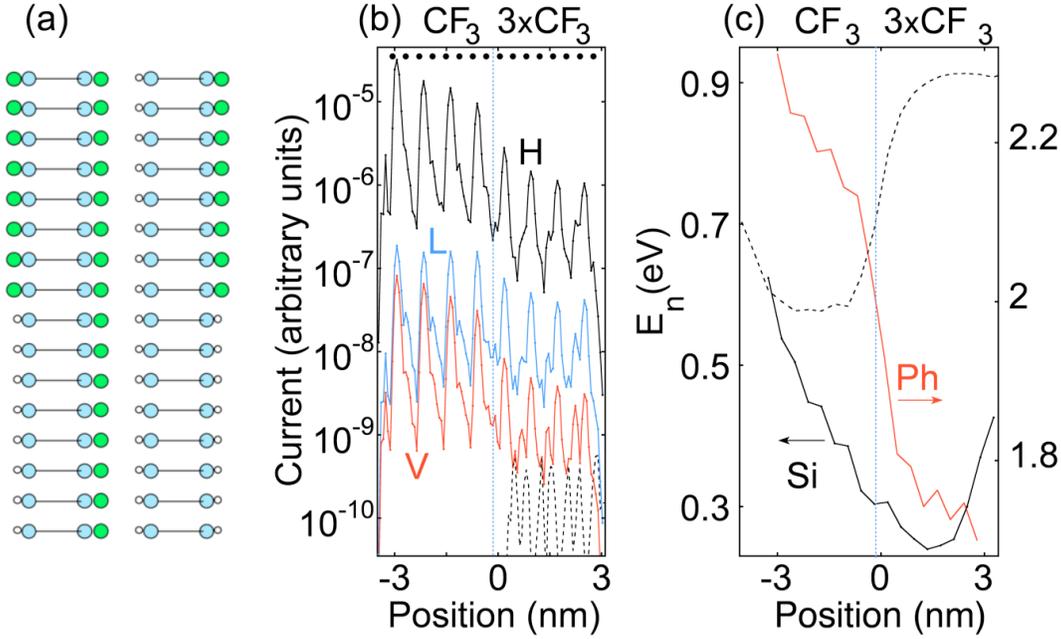}
\caption{\label{Fig_13th}Color online.
(a) Schematic top view (not to scale) of 
model single-triple CF$_3$--styrene structure. Molecules are green,
surface Si atoms blue, H atoms white.  
The single CF$_3$--styrene row consists of 8 molecules, the
triple row of 24 molecules. All molecules are T-tethered
to Si; see Fig.\ref{Fig_4struct}(b).
(b) Plots V, L and H are calculated current profiles 
for an STM tip trajectory at constant height along
center row of CF$_3$--styrene molecules in (a)
for negative substrate 
bias. 
Plot V: Low bias; tip Fermi level near highest Si valence
band states. Plot L: Stronger bias but tip Fermi level above
CF$_3$--styrene HOMO energies. \cite{lowerheight} 
Plot H: Still stronger bias; tip Fermi level well within
the HOMO energy band of the {\em single} CF$_3$--styrene row.
As in experiment, contrast between triple and
single CF$_3$--styrene rows in blue  profile L 
is much weaker than in red
profile V. Dashed black curve:  Calculated current profile
along short CF$_3$--styrene row on left in (a) at
same tip height and bias as plot V. 
Black bullets locate C atoms of CF$_3$ groups
in the long CF$_3$--styrene row.  (c) Black
solid curve (Si): Electrostatic electronic energy shifts $E_n$ at Si atoms to
which molecules of the long CF$_3$--styrene row bond. Triple (single)
rows are right (left) of vertical dotted line. 
Red curve (Ph): 
Average of $E_n$ over the six carbon atoms of the benzene ring
of each molecule of the long CF$_3$--styrene row (scale on right axis).
Black dashed curve shows for comparison $E_n$ at Si atoms to
which molecules of a {\em single}-row heterostructure of 
10 CF$_3$--styrene molecules
(on the left) and 10 OCH$_3$--styrene molecules (on the right) bond,
computed for a silicon substrate cluster having the same cross-section.}
\end{figure*}

Representative results of our calculations are presented in Fig. \ref{Fig_13th}.
The molecular geometry studied is shown schematically in 
Fig.\ref{Fig_13th}(a). In the model considered 
the long CF$_3$--styrene line is located between the two short ones to minimise
sensitivity to the Si cluster edges.  All of the molecules are assumed
to be tethered to the silicon in the T orientation defined in Section \ref{Geom}
and shown in Fig. \ref{Fig_4struct}(b).
Fig.\ref{Fig_13th}(b) shows the simulated constant height current 
along the long (central) CF$_3$-styrene line. At low negative 
substrate bias (for which
the STM tip Fermi level is close to the silicon valence 
band edge), the calculated 
current (curve V) drops by a factor of $\sim$26 from the tallest current 
peak near the
left hand end of the single CF$_3$-styrene line to the tallest
peak near the right hand end of the central line of the triple 
CF$_3$-styrene structure.
The origin of this effect is seen in the solid black curve labeled
``Si'' in Fig.\ref{Fig_13th}(c):  Dipole fields associated with the 
CF$_3$-styrene molecules 
lower the Si orbital energies below the triple
CF$_3$-styrene by
$\sim$0.2eV more than under the single file CF$_3$-styrene.
At low bias this reduces the silicon electronic density of states
near the silicon surface at the Fermi level of the STM tip 
more under the  triple CF$_3$-styrene line than under the
single CF$_3$-styrene line, resulting in the weaker tip current
through the former. As the magnitude of the bias increases
eventually the tip Fermi level is lowered further by an amount 
substantially exceeding the differential electrostatic energy
shift between the Si orbitals under the triple and single 
CF$_3$-styrene lines and thus the contrast between the
STM tip currents through the  triple and single 
CF$_3$-styrene lines becomes weaker as is seen
in curve L of Fig.\ref{Fig_13th}(b) for which the tip
Fermi level is 0.5eV lower than for curve V.

This behavior found in the simulations
(lower current through the triple CF$_3$-styrene line
than through the single CF$_3$-styrene line 
at low bias and the reduction
in this contrast with increasing magnitude of the
bias) is qualitatively similar to that seen experimentally 
in Fig. \ref{Fig_3expt}(e). However, quantitatively the
suppression of the current through the
triple CF$_3$-styrene line 
at  low bias is more pronounced in the experiment
(Fig. \ref{Fig_3expt}(e))
than in the theoretical curve V of Fig.\ref{Fig_13th}(b): 
A 0.2-0.3nm difference in tip height
at constant current (as  between the triple and single 
CF$_3$-styrene lines in Fig. \ref{Fig_3expt}(e) at low bias)
normally corresponds to approximately
a 2-3 orders of magnitude change in current at constant
tip height in STM experiments. 

Part of this difference between the simulation and experiment
may be due to the unknown microscopic details of the STM tip,
which are known to influence the height contrast between
CF$_3$-styrene and OCH$_3$-styrene lines 
as is discussed in 
Section \ref{Expt} and can be seen in Fig.\ref{Fig_2expt}.
However, the limited size of the model system for which
the present simulations could be carried out\cite{size} is clearly
responsible for at least a part of the difference between the
experimental and theoretical findings:  While the downward
electrostatic shift of the Si orbital energies (as indicated
by the solid black curve in Fig.\ref{Fig_13th}(c)) is clearly
larger under the triple than the single CF$_3$-styrene line,
it is evident that the 8 molecule {\em single} CF$_3$-styrene line
modeled is not long enough for the electrostatic potential
in the silicon under the  single CF$_3$-styrene line to 
reach a plateau with increasing distance from the 
junction with the triple  CF$_3$-styrene line, i.e., the
electrostatic influence of the triple row of molecular
dipoles of the triple CF$_3$-styrene line
is clearly being strongly felt under much of  the single
CF$_3$-styrene line as well, resulting in diminished 
contrast in the calculated current between the single 
and triple CF$_3$-styrene lines  relative to that which
may be expected in larger model systems. In this regard
it is instructive to compare the electrostatic profile in the
silicon under the single-triple CF$_3$-styrene  
heterostructure with that for a somewhat 
longer {\em single} file 
CF$_3$-styrene/OCH$_3$-styrene heterostructure
shown by the dashed black curve in Fig.\ref{Fig_13th}(c):
For the latter heterostructure the potential profile
in the silicon under the CF$_3$-styrene line (which
is on the left) does show a pronounced plateau where
the value of the electrostatic potential is close to that
for the single-triple CF$_3$-styrene heterostructure
under the {\em end} of the single CF$_3$-styrene line
that is {\em furthest} from the triple CF$_3$-styrene line.

In comparing theory with experiment it is
noted that there is also
contrast
between the tip currents calculated for tip positions
over the three molecular rows within the
triple row structure: The calculated current 
for the tip over the leftmost molecular row in 
Fig.\ref{Fig_13th}(a) for the same tip height and
bias as for plot V of Fig.\ref{Fig_13th}(b) is 
shown by the dashed curve at the bottom of
Fig.\ref{Fig_13th}(b): The current over the left
molecular row is 
 smaller  than over the center row of the triple
structure (plot V). The calculated current over 
the rightmost
row of Fig.\ref{Fig_13th}(a) (not shown) is 
intermediate between that for the left and
center rows. These results suggest that stronger
contrast between the single CF$_3$-styrene line
and parts of the triple CF$_3$-styrene than
is seen in plot V of Fig.\ref{Fig_13th}(b) may
be possible theoretically.
However, calculations for larger
silicon clusters than were feasible in this 
study are required to eliminate possible
cluster edge effects and develop a 
better understanding of this issue.

If the magnitude of the negative substrate
bias is increased still further, well beyond that for current
plot L in Fig.\ref{Fig_13th}(b), the STM current through the
single
CF$_3$-styrene line is predicted by the present 
simulation to begin to increase more rapidly
than that through the central line of the triple
CF$_3$-styrene structure. This is seen in curve H
of Fig.\ref{Fig_13th}(b) where the current through
the single
CF$_3$-styrene line is again much larger
than that through the triple
CF$_3$-styrene line. This effect can be understood
by considering the red curve (Ph) in Fig.\ref{Fig_13th}(c):
This shows 
the average of $E_n$ over the six carbon atoms of the benzene ring
of each molecule of the long CF$_3$-styrene row. 
This is higher for the single
CF$_3$-styrene line than for the  triple
CF$_3$-styrene line. Therefore the molecular HOMO
levels for the molecules in the single CF$_3$-styrene line
are higher in energy than those for molecules in the center row
of the triple CF$_3$-styrene line. Consequently the
molecular HOMO
levels for the molecules in the single CF$_3$-styrene line
are crossed by the Fermi level of the STM tip at lower
magnitudes of the negative substrate bias than for 
molecules in the  triple CF$_3$-styrene line which
results in the renewed enhancement of the current
through the single CF$_3$-styrene line relative to
the triple CF$_3$-styrene line seen in curve H
of Fig.\ref{Fig_13th}(b). No recovery of stronger
contrast between the single and triple CF$_3$-styrene lines
at stronger negative substrate bias is seen experimentally
in Fig. \ref{Fig_3expt}(e) indicating that  even for the 
strongest negative substrate bias values realized in these
experiments the STM tip Fermi level remains well above
the  CF$_3$-styrene HOMO levels even for the molecules
in the single CF$_3$-styrene line. This is consistent with
our experimental findings for the 
CF$_3$-styrene/OCH$_3$-styrene heterostructures
where no acceleration of the increase of the apparent height
of the CF$_3$-styrene with bias at strong negative substrate
bias (that would be  indicative of the STM tip Fermi level crossing the 
CF$_3$-styrene HOMO) was observed.

\section{Discussion}
\label{conclusions}

The present experimental and theoretical work has  begun to explore 
how electric fields emanating from molecules influence electrical
conduction through other molecules in their close vicinity 
and in the underlying substrate.

In this work we have employed 
density functional calculations only to estimate the electrostatic potential
throughout the system in its electronic ground state, an application
of density functional theory that is believed to be soundly based at
the fundamental level. The phenomena that we report on here can be
understood {\em qualitatively} by considering just the results of these electrostatic
calculations as summarized in Fig. \ref{Fig_7pots} and \ref{Fig_13th}(c). 
Transport 
calculations are however necessary to translate the electrostatics into current
profiles that can be compared with the experimental STM
data. 
As is discussed in Section \ref{Formalism}, 
transport calculations that incorporate electronic structures obtained
entirely from density functional theory, although popular, are
not soundly based and are often misleading, 
especially for molecules on semiconductors.
We therefore base our transport calculations instead on 
electronic structures obtained from
semi-empirical tight binding models modified to include
the density functional theory-based ground state 
electrostatic potentials.

Our theoretical modeling indicates the  
observed low bias filled-state 
current enhancement in the interfacial OCH3-styrene 
molecules to be due to the
collective effect of the electric dipole fields generated by the OCH$_3$ 
groups of the OCH$_3$--styrene chain when the OCH$_3$ 
groups are aligned preferentially so that their carbon atoms are
further from the heterojunction than their oxygen atoms, as they are,
for example, in Fig. \ref{Fig_8th}(a).
It is plausible that structures of this sort are prevalent in these systems 
since some of them should be favored for molecules subject to a weak
net mutual steric repulsion as is discussed  Section \ref{Structure}. 
However the resolution of our experimental STM data which was taken
at room temperature is not sufficient to
determine independently whether or not the molecular chains
have structural order of this sort. Furthermore it is apparent from a 
comparison of our experimental data and theoretical results
that  the CF$_3$ groups in the CF$_3$--styrene molecules must be
spinning rapidly enough on their axes that only an average over
many conformations of these molecules is being observed in the
STM images. Similarly it seems possible that many conformations 
of the OCH$_3$--styrene molecules are being averaged over
rapidly in the STM imaging and that the above alignment of the OCH$_3$ 
groups is simply more commonly present than other structures, such as 
that in Fig. \ref{Fig_8th}(b)
for which the apparent height enhancement should be near the
end of the OCH$_3$--styrene chain that is remote from the junction.

The behavior of the imaging height enhancement that we observe
experimentally with increasing negative substrate bias is also consistent
with the predictions of our theoretical model of this phenomenon.

Under positive substrate bias simulations indicate that height 
enhancement near the junction should also occur but should 
in this case be confined to only a single molecule. The location 
of the predicted height enhancement under positive substrate 
bias is sensitive to the details of the local molecular 
geometry in its vicinity that are expected to fluctuate rapidly at 
room temperature. 
Thus definitive observation of this effect will 
most likely require cryogenic 
experimental work.

The work reported here has succeeded in identifying transport 
phenomena occurring at molecular length scales that can 
reasonably be attributed to electric fields emanating from polar 
molecules impinging on other nearby molecules and on the 
underlying substrate, and in developing some insights into the 
detailed physical mechanisms that may be involved. As has been
discussed in Section \ref{Intro} these phenomena may ultimately 
find device applications, such as molecular switches 
in which the current through a molecule is switched by 
conformational changes in another nearby molecule
and/or devices in which conduction through a semiconductor
substrate is controlled with the help of molecules at its
surface.
  
However further experimental and theoretical work is
needed in order to better understand and ultimately
control such phenomena.

\section*{Acknowledgments}
This research was supported by the Canadian Institute
for Advanced Research, NSERC, iCORE and the NRC. 
PGP wishes to thank NRC-INMS for support.
Some numerical
computations presented in this work were performed on
WestGrid computing resources, which are funded in part by
the Canada Foundation for Innovation, Alberta Innovation
and Science, BC Advanced Education, and the participating
research institutions. WestGrid equipment is provided by
IBM, Hewlett Packard, and SGI. We
have benefited from discussions with G. DiLabio and from the  
technical expertise of D. J. 
Moffatt and M. Cloutier.

%
%
%

%
%
%
%

%
%

%

%

%

%

%

\begin{thebibliography}{100}
\bibitem{caveat}Certain substituents have been found to hinder the line 
growth process, and on occasion
successful heterowire formation has been observed to depend on deposition 
sequence.
\bibitem{Gaussian} The Gaussian03 package with the B3PW91
density functional and Lanl2DZ basis set was used.
\bibitem{temp}The temperature $T$ in the Fermi functions
was set to 0 in the calculations reported here.
\bibitem{GaussianUFF}The numerical implementation of the 
Gaussian98 package was used.
\bibitem{lobiasexpt}In our STM experiments imaging of the CF$_3$--styrene/OCH$_3$--styrene 
molecular heterowires in this very low bias regime was rarely possible,
as is discussed in Section \ref{Level Ordering}.
\bibitem{size} We include 423 Si atoms and 936 other atoms in our
density functional calculations of the electrostatic potentials
in the single-triple row CF$_3$--styrene heterostructure on 
silicon.
\end{thebibliography}
\end{document}